\PassOptionsToPackage{square,numbers,sort&compress}{natbib}
\documentclass[preprint,12pt]{elsarticle}
\usepackage{multirow} 
\usepackage[a4paper,top=2cm,bottom=2cm,left=2.5cm,right=2.5cm,marginparwidth=1.75cm]{geometry}
\usepackage{amssymb}
\usepackage{caption}
\usepackage{subfigure}
\usepackage{float}
\usepackage{hyperref}
\usepackage{amssymb}
\hypersetup{hidelinks,
	colorlinks=true,
	allcolors=black,
	pdfstartview=Fit,
	breaklinks=true}
\usepackage{amsmath}
\usepackage{graphicx}
\usepackage{pdfpages}
\usepackage{upgreek}
\usepackage[toc,page]{appendix}
\usepackage{url}
\journal{Elsevier}

\begin{document}
\begin{frontmatter}

\title{Effect of fiber curvature on gas diffusion layer two-phase dynamics of a  proton exchange membrane fuel cell}
\author[inst1]{Danan Yang \corref{mycorrespondingauthor}}
\cortext[mycorrespondingauthor]{Corresponding author}
\ead{danan.yang@energy.lth.se}
\author[inst1]{Himani Garg}
\affiliation[inst1]{organization={Department of Energy Sciences, Faculty of Engineering, Lund University},
            addressline={P.O. Box 118}, 
            postcode={Lund, SE-221 00}, 
            country={Sweden}}
\author[inst1]{Martin Andersson}

\begin{abstract}

The dynamics of two-phase flow within the cathode of a proton exchange membrane fuel cell, particularly in Gas Diffusion Layers (GDLs) with varying fiber curvatures, remain underexplored. Using a periodic surface model, we stochastically reconstruct three GDL types with different fiber curvatures, incorporating vital parameters derived from a physical GDL. Considering the randomness in reconstruction, the structure generation process is iterated four times for each GDL type, enabling an ensemble average analysis. Pore network models are adopted to reveal disparities in these GDL porous structures. The subsequent two-phase simulations are conducted to explore liquid transport through these GDLs and interfaces to assembled gas channels. Time-varying GDL total, local water saturation, and capillary pressure are investigated. Results show stochastic reconstructions exhibit similar frequency peak ranges in pore and throat diameters, and coordination numbers, but diverge from the physical GDL. Bigger fiber curvature tends to enhance pore network connectivity by increasing smaller pores,  leading to heightened water saturation and capillary pressure. Straight-fiber GDLs, compared to curved-fiber GDLs, show greater potential proximity to the physical GDL in terms of overall water saturation and capillary pressure but are also accompanied by increased uncertainty. Despite similar layer porosity, water saturation in the same layer of all samples differs increasingly from the inlet to the outlet. Water breakthrough and detachment near the GDL can induce significant water saturation instability at the GDL and gas channel interface. Detached droplets in gas channels connected with straight-fiber GDLs exhibit larger sizes and slower movement than those in channels assembled with curved-fiber GDLs. These findings can be utilized in future GDL design and optimization.
\end{abstract}

\begin{graphicalabstract}
\begin{figure}[H]
\centering
    \centering
    \includegraphics[width = 0.8\textwidth]{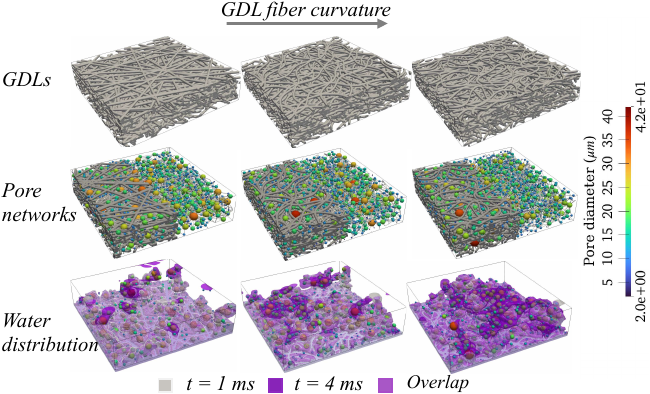}
\end{figure}
\end{graphicalabstract}

\begin{highlights}
\item[-] {Curved and straight fiber GDL reconstructions with similar bulk and layer porosity.}
\item[-] {Larger fiber curvature enhances the connectivity of the pore network and increases the number of smaller pores.}
\item[-] {GDL water saturation and capillary pressure increase with fiber curvature.}
\item[-] {Intimate correlation between GDL pore network and two-phase flow dynamics is studied.}
\item[-] {Breakthrough water instability is related to the droplet detachment location.}
\end{highlights}

\end{frontmatter}

\section{Introduction}
Proton exchange membrane fuel cells are anticipated to utilize hydrogen energy to generate electricity based on the electrochemical reaction with oxygen with the promise of high efficiency and pollution-free operation \cite{wang2021review}. Nevertheless, effective management of the simultaneously produced heat and water is essential to ensure optimal performance \cite{liu2022review}. In the typical operation temperature range of 60-80 $^{\circ}$C \cite{qu2022proton}, the generated water vapor within the cathode catalyst layer tends to undergo condensation. The liquid water then flows through the Gas Diffusion Layer (GDL) before reaching the Gas Channels (GCs). Excessive liquid water accumulation inside the GDL will impede the diffusion of reactive gas from the GCs, particularly under the condition of high current density, in turn, leading to a performance decrease and even irreversible damage \cite{merida2006characterisation}. The mechanisms of gas and liquid water transport inside the GDL remain challenging due to the intricate nature of small scales and porous media \cite{okonkwo2021review}. Therefore, understanding the interaction between liquid water movement and GDL microstructure as well as further water behavior in the GDL/GC interface and GC is crucial to optimize water management.

Commercially available GDLs are predominantly carbon-based porous mediums, which are manufactured in different types, e.g., carbon paper, carbon felt, carbon cloth, and carbon foam \cite{athanasaki2023gas}. All of these GDLs exhibit distinctions in terms of fiber curvature, fiber diameter, porosity size distribution, binder, polytetrafluoroethylene treatment, thickness, compression size, and so on. To investigate these GDL features and the inside two-phase dynamics, both experimental and numerical studies have been conducted. Experimental approaches generally rely on high-resolution (about 1.5-10 $\mu m$) visualization techniques, i.e., fluorescence microscopy \cite{hasanpour2018woven}, X-ray computed tomography \cite{lee2017investigating, muirhead2018liquid, mularczyk2020droplet}, and neutron radiography \cite{owejan2014oxygen, maier2020mass}. Hasanpour et al. \cite{hasanpour2018woven} studied the water behavior within woven GDLs using fluorescent microscopy, and the results show that applying a fluorinated ethylene propylene coating to the GDL can benefit water removal and higher thermal conductivity but reduce electrical conductivity. The interaction of water removal and oxygen transport inside the GDLs was investigated using synchrotron X-ray radiography in different conditions of cathode inlet relative humidity and operating current density \cite{muirhead2018liquid}. A combined neutron imaging and X-ray computed tomography was developed by Maier et al. \cite{maier2020mass}, to study the liquid-gas transport in the GDL of a proton exchange membrane electrolyzer.

In contrast to experimental techniques that demand expensive high-resolution machinery and advanced image-processing approaches, along with the real GDL frameworks, numerical methods provide cost-effective and flexible options. There are three popular numerical methods including the lattice Boltzmann method \cite{jeon2015effect, zhu2021pore, yang2022lattice, zhang2022study},  pore network model \cite{straubhaar2016pore, zapardiel2022modeling}, and volume of fluid method \cite{zhou2019two, bao2021transport, jiao2021vapor, jiao2022investigations, yang2023numerical}. These methods enable the exploration of essential mechanisms that might remain elusive or limiting in experiments. Moreover, they also provide the advantage of flexible control over geometric and operational variables to seek optimal GDL design and reasonable operation. Pore network models are usually based on simplified GDL geometries, such as regular pore and throat structures. Therefore, they have a cheaper computational burden and better capacity in large-scale simulations. Both lattice Boltzmann methods and volume of fluid are adopted for complex GDL structures. Nevertheless, the majority of lattice Boltzmann methods are employed for simulation within the two-dimensional GDL domain, with limited application in three-dimensional scales. The computational cost of the volume of fluid simulations is notably high, particularly in three-dimensional scenarios (from days to months), still,  
it can better guarantee the real GDL features with an appropriate mesh resolution. Straubhaar et al. \cite{straubhaar2016pore}  and Jiao et al. \cite{jiao2021vapor} have studied condensation within the GDLs using the pore network model and volume of fluid, respectively. Furthermore, the impact of compression on liquid water transportation in the GDL is investigated with lattice Boltzmann methods \cite{jeon2015effect, zhu2021pore} and volume of fluid \cite{zhou2019two, bao2021transport}.

During the numerical investigation, the GDL structures usually need to be reconstructed either using an image-based method based on image sequence from tomography scanning and electron microscope \cite{fluckiger2011investigation, satjaritanun2017micro, totzke2014three, zenyuk2016gas} or using a geometry-based stochastic method \cite{xiao2019solid, shi2022compressive, yang2023numerical}. The stochastic methods have attracted enough attention to GDL reconstruction with straight carbon fiber, e.g., the Toray-type GDLs. In a GDL with constant bulk porosity, increasing the fiber curvature is likely to reduce the number of fibers, which shows great potential to change pore structures. Besides, some commercialized GDLs like Fredeunberg GDLs are manufactured with curved fibers \cite{zenyuk2016gas, zhang2021microstructure, chuang2020interactive, tranter2020pore}. However, the reconstructions of carbon cloth and carbon felt consisting of curved carbon fibers are still rare. In contrast to the stochastic straight-fiber GDL reconstruction, where a fiber orientation is determined by two random points \cite{yang2023numerical}, curved-fiber GDLs are reconstructed based on a curved line \cite{gaiselmann2013stochastic} or periodic surface \cite{didari2017modeling}. Gaiselmann et al. \cite{gaiselmann2013stochastic} utilized a random walk algorithm based on multivariate time series to randomly generate curved track lines by inputting extracted fiber features from synchrotron tomography. Didari et al. \cite{didari2017modeling} improved the periodic surface model proposed by Wang et al. \cite{wang2007periodic} to reconstruct a curved-fiber GDL and showed similar properties with the actual carbon felt Freudenberg H2315 GDL in terms of the cumulative density function of its pore size distribution. In addition, the tortuosity, in-plane permeability, through-plane permeability, and chord length are also compared.

To the best knowledge of the authors, most GDL studies only pay attention to straight-fiber GDLs, and the effect of fiber curvature on water behavior has not yet been studied. The pore and throat spatial distribution of the GDLs with different fiber curvature is rarely studied by the pore network. Therefore, based on curved-fiber and straight-fiber GDL configurations reconstructed by using an improved periodic surface model, the GDL fiber curvature influence on the inner GDL water distribution is numerically investigated for the first time using the volume of fluid method in OpenFOAM 7.0. The water behavior within the GDL and breakthrough interface is investigated. Moreover, the relationship between the water dynamics and the pore networks is analyzed. Note the purpose of this work is not only to try to mimic the water behavior inside a real GDL using stochastically reconstructed GDLs with different fiber curvatures but also to study the effect of different fiber curvatures on GDL pore structure and water behavior.

\section{Methods} 

\subsection{Computational domain}
\begin{figure}[H]
    \centering
    \includegraphics[width=0.7\textwidth]{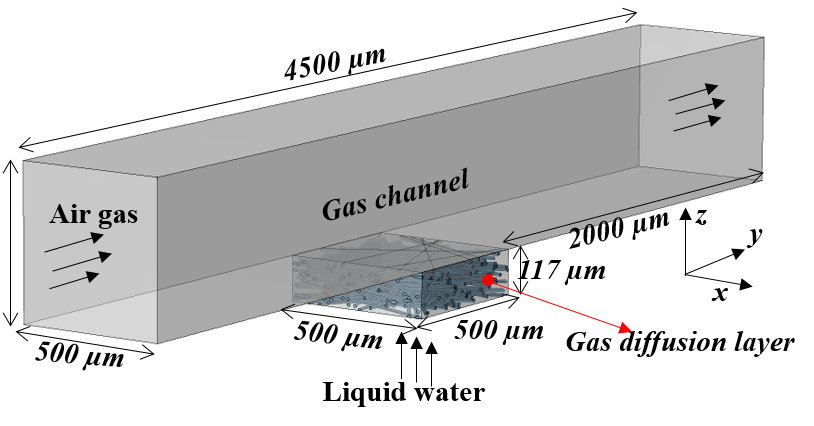}
    \caption{A "T-shaped" configuration for the numerical simulation, including a long GC above and a smaller porous GDL below the middle of GC.}
    \label{fig:geometry}
\end{figure}

In this work, a 'T-shaped' computational geometry is considered, consisting of a fibrous GDL and an assembled GC, as shown in Fig. \ref{fig:geometry}. The long GC is exclusively supplied with air, while the entire bottom of the GDL serves as the liquid inlet. Previous studies that only consider a single GDL have neglected the effects of airflow on the water behavior close to the GDL top region \cite{zhou2019two, bao2021transport}. Therefore, the assembled GC is used to conduct more realistic boundary conditions. For example, it should meet the natural development of airflow and the water flow behavior along the airflow direction. Due to the significant computational load, the reconstructed GDL size is smaller than that of GC. The dimensions of each component are labeled in the geometry.  

\subsection{Stochastic GDL reconstruction}
\begin{figure}[H]
    \centering
    \includegraphics[width=1\textwidth]{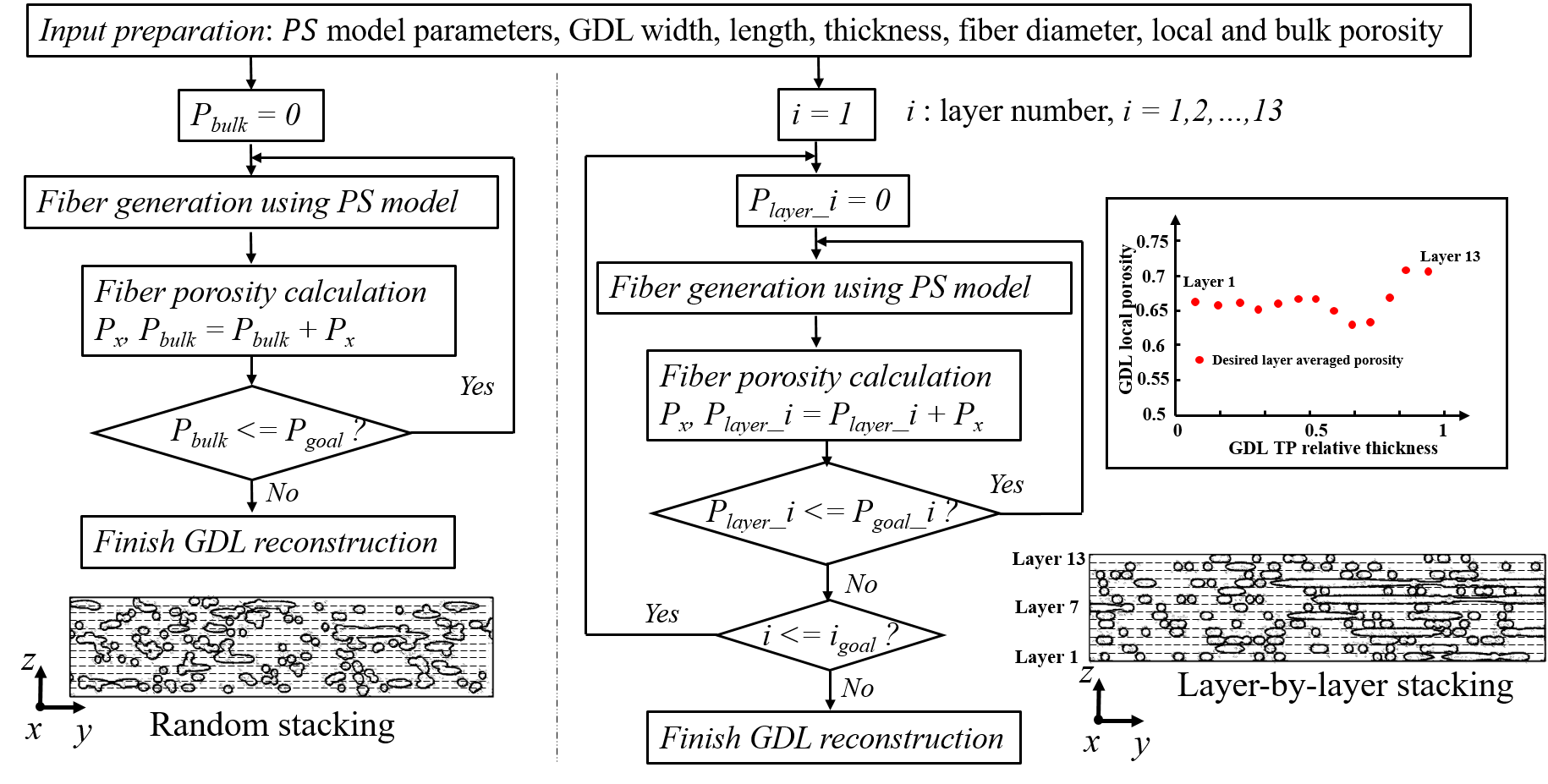}
    \caption{Stochastic GDL reconstruction schemes based on a rod periodic surface model ($PS$: periodic surface; $P_x$: the $x^{th}$ fiber porosity; $P_{bulk}$: bulk porosity; $P_{goal}$: desired porosity; $i$: layer number, 1,2,...,13.).}
    \label{fig:reconstruction}
\end{figure}

\begin{figure}[H]
    \centering
    \subfigure[]{\includegraphics[width=0.48\textwidth]{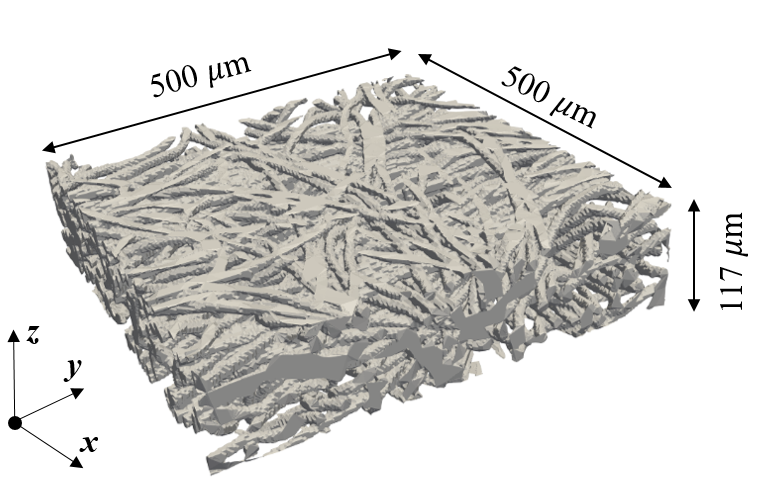}}
    \subfigure[]{\includegraphics[width=0.48\textwidth]{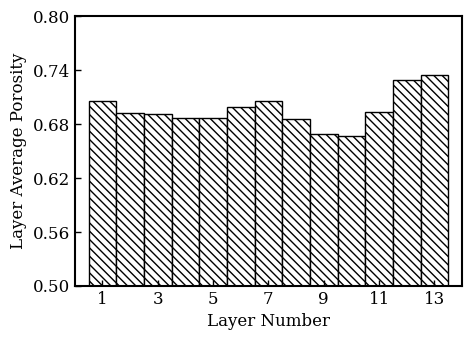}}
    \caption{(a) A physical Fredeunberg H2315 GDL sample, measuring 500 µm in length, 500 µm in width, and 117 µm in thickness; (b) The through-plane layer porosity distribution of 13 layers, with Layer 1 near the GDL bottom and Layer 13 adjacent to the GDL/gas channel (GC) interface}
    \label{fig:ExpLocalLayer}
\end{figure}

To obtain the required GDL structure topology and study the effects of different fiber curvature, a stochastic GDL reconstruction is chosen due to its efficiency and flexibility in adjusting parameter values. The proposed GDL reconstruction process is shown in Fig. \ref{fig:reconstruction}. To make the GDL reconstruction more realistic, some vital parameters are extracted from a physical GDL structure, Freudenberg H2315 (shared by Forschungszentrum Jülich GmbH), as illustrated in Fig. \ref{fig:ExpLocalLayer}(a). These parameters include the desired domain size (500 $\mu m$ $\times$ 500 $\mu m$ $\times$ 117 $\mu m$), fiber diameter (9 $\mu m$), through-plane local porosity, and bulk porosity (around 0.7). The through-plane local layer porosity is shown in Fig. \ref{fig:ExpLocalLayer}(b). Furthermore, the presented H2315 GDL topology reveals the presence of both straight and curved fibers.

GDL fibers are generated using the periodic surface model proposed by Wang et al. \cite{wang2007periodic, huang2015generalized}. There are various periodic surface models for different shapes of structure surface, such as sphere, membrane, and mesh shape. In the present work, a rod periodic surface model $f(\mathbf{r})$ is selected to reconstruct the GDL fibers, as shown below,  
\begin{equation}
\begin{aligned}
     f(\mathbf{r}) &= 4\cos(2 \pi (\mathbf{RT}P_1)^T\mathbf{r} + b\cos(2 \pi f_r(\mathbf{RT}Q_1)^T\mathbf{r} )) +4\cos(2 \pi (\mathbf{RT}P_2)^T\mathbf{r}) + \\
     &\phantom{=} 4\cos(2 \pi (\mathbf{RT}P_2)^T\mathbf{r}) + 3\cos(2 \pi (\mathbf{RT}P_3)^T\mathbf{r}) - 4\cos(\pi (1-S_r)) + 1
\end{aligned}
\end{equation}

Where, $\mathbf{r}$ = $[x,y,z,1]$ is the location vector within the unit space $\mathbb{R}^3 \in [0,1]^3$. To suit different length scales, a scaling factor $S_r$ is introduced to scale the generated structure to the desired dimension. Specifically, $S_r$ is the ratio of fiber diameter and the longest dimension of the expected GDL, which are 9 $\mu m$ and 500 $\mu m$, respectively. $[P_1, P_2, P_3, P_4]$ and $Q_1$ are fiber display orientation and deforming orientation matrices. $\mathbf{R}$ and $\mathbf{T}$ are translation and rotation matrices, which are the functions of $(\alpha, \theta, w)$ and $(t_1, t_2, t_3)$. $\alpha$, $\theta$, and $w$ are axis rotation angle align $x$,$ y$, and $z$ axis, respectively. $t_1$, $t_2$, and $t_3$ are translation scalar align $x$-$y$, $y$-$z$, and $x$-$z$ planes. Moreover, $b$ and $f_r$ are used to control the fiber curvature magnitude and wave frequency, respectively.
A more detailed discussion can be found in previous works \cite{yang2023numerical}.

In addition, GDL reconstruction depends on a specific fiber stacking strategy for the generated fibers. Random stacking is commonly used in previous GDL reconstructions \cite{didari2017modeling, jiao2022investigations, zhou2019two, bao2021transport}. In this study, a layer-by-layer stacking method is adopted to maintain a desired layer porosity distribution along the through-plane direction during reconstructions. Moreover, the bulk porosity is difficult to keep constant in GDL reconstructions, thus a maximum tolerance of 1.5 \% is given to it.

\subsection{Computational methodology}
In this work, the gas and liquid flow in the whole region is considered to be incompressible and laminar. Temperature is assumed to be constant at 60 $^\circ$C within the specified simulation domain. In order to characterize the dynamics of two-phase flow, the \textit{interFoam} solver, utilizing the volume of fluid methods within OpenFOAM 7.0, is employed for the resolution of mass and momentum equations for gas-liquid mixtures, alongside an advection equation governing fluid fraction \cite{hirt1981volume}. The equations are expressed as,   

Mixture mass and momentum conservation equations:
\begin{equation}
\nabla\cdot\mathbf{u} = 0
\end{equation}
\begin{equation}\label{Momentum conservation equation}
\rho\frac{\partial{\mathbf{u}}}{{\partial t}} + \rho\nabla\cdot\left(\mathbf{u}\mathbf{u}\right) = \nabla\cdot\left[\mu\left(\nabla\mathbf{u}+\nabla\mathbf{u}^T\right)\right] -\nabla p + \rho\mathbf{g}+ {\mathbf{f}_{\sigma}}
\end{equation}

Fluid-fraction advection equation:
\begin{equation}
\frac{\partial\alpha}{\partial t} + \nabla\cdot\left(\alpha\mathbf{u}\right) = 0 
\end{equation}

In a fluid system characterized by two phases, $\alpha$ represents the volume fraction of one phase within the specified control volume, with 1-$\alpha$ denoting the volume fraction of the other phase. $\mathbf{u}$ and $p$ are the velocity and pressure, respectively. $\mu$ and $\rho$ are the fluid mixture viscosity and density, which are weighted-averaged by $\alpha$ based on two-phase corresponding properties. $\mathbf{f}_{\sigma}$ is surface tension force, and $\mathbf{g}$ is gravity acceleration. More details of the governing equations can be found in previous research \cite{hirt1981volume}. 

\subsection{Boundary conditions}
In addition, boundary conditions are applied to different solid surfaces. Uniform velocity boundary conditions are enforced at GC inlet air and GDL inlet liquid with 10 m/s and 0.02 m/s, respectively. Bigger liquid inlet velocity value of 0.054 \cite{zhang2018three} and 0.1 m/s \cite{niblett2020two, liu2023numerical} were utilized. It should be mentioned that liquid inlet velocity has been scaled more than 1000 times compared with that in the real operation, around 2E-5 m/s \cite{niblett2020two, sarkezi2023lattice}. An ex-situ experiment has shown that the liquid breakthrough takes several minutes \cite{lu2010water}. Furthermore, the time steps in such micro-scale simulations with a grid resolution around several micros are typically small. Therefore, it will take more than 1.5E5 core-hours computational time to see the liquid breakthrough if employing such a physical GDL liquid invasion velocity. Fortunately, it was found that scaling up the liquid velocity causes little change in GDL capillary fingering dominated flow and dramatically accelerates the simulation to observe longer water behavior \cite{sakaida2017large}. At the GC outlet, a zero gradient condition for the velocity and liquid volume fraction, together with a total pressure condition, is applied. At the GDL side walls, a symmetry boundary condition is adopted to mimic the influence of the two-phase flow surrounding the GDL region. For the GC and GDL surfaces, zero flux and no-slip boundary conditions are enforced.
Moreover, the surface wettability is determined by varying contact angles, with the GDL surface and GC bottom surface exhibiting a contact angle of 150 $^\circ$, while the remaining three GC surfaces have a contact angle of 45 $^\circ$. Considering the trade-off between optimal computational speed and simulation accuracy, all the equations are discretized by mixed one-order and two-order accuracy. Besides, an adjustable time step strategy is employed by controlling the Courant number below 0.5. In addition, the simulation timestep is observed to stabilize around 4E-8 s. Moreover, the liquid water and gas transport property values, e.g.,  viscosity, surface tension, and density, are utilized considering the operation temperature of 60 $^\circ$C. Note that all simulations in this study are performed on parallel computing clusters, totally utilizing over 105,000 CPU hours. Each case requires around 48 hours of computation using 160 cores.

\subsection{Capillary pressure calculation}
Liquid water transport inside a GDL is dominated by the capillary fingering process \cite{xu2021temperature}. The related capillary pressure $P_c$ is defined by,
 
\begin{equation}\label{Eq1}
    P_c = P_l - P_g 
\end{equation}
Where, while $P_l$ and $P_g$ denote the liquid and gas pressure at the respective sides of the two-phase interface. To obtain the liquid and gas pressure, the fluid fraction $\alpha$ in the volume of fluid simulation is utilized to identify two phase domains. Theoretically, $\alpha = 1$ represents the volume occupied by the liquid phase, whereas $\alpha = 0$ donates the volume filled by the gas phase.  The two-phase interface has a step variation in volume fraction, but there is usually a thin transition region where $0 < \alpha < 1$ between the two phases in numerical simulation, and $\alpha = 0.5$ is widely considered as the interface between two phases. The pressure distribution within each phase remains relatively uniform. Therefore, we calculate the averaged pressure difference across two specific interfaces within the interfacial region. One interface corresponds to an isosurface $\mathbf{\Omega_l}$ with a liquid fraction of 0.51 and another is an isosurface $\mathbf{\Omega_g}$ with a liquid fraction of 0.49, which contains the pressure of the gas phase, a similar method can be found in \cite{rabbani2016effects}. Mathematically, the capillary pressure is calculated according to the following equation.
\begin{equation}
    P^{interface}_c = \frac{\sum_{i = 1}^N P_{l,i}}{N} - \frac{\sum_{j = 1}^M P_{g,j}}{M}
\end{equation}
Here, $P_{l, i}$ and $P_{g, j}$ are the liquid pressure at the $i_{th}$ point in $\mathbf{\Omega_l}$ and gas pressure at the $j_{th}$ point in $\mathbf{\Omega_g}$. N and M separately represent the number of grid points in $\mathbf{\Omega_l}$ and $\mathbf{\Omega_g}$.

\subsection{Pore network extraction}
Young-Laplace equation is also used to estimate the capillary pressure, i.e., $P_c = 2\sigma cos(\theta)/r$. Here, $\sigma$ represents the surface tension force. $\theta$ is the contact angle and $r$ is the pore radius. Under conditions of a consistent contact angle and surface tension, capillary pressure demonstrates an inversely proportional relationship with pore size. Therefore, the pore network serves as a crucial intermediary connecting the complex GDL structure with its inside water dynamics, which illustrates the pore and throat locations and connections within the GDL. Besides,  before conducting relatively expensive two-phase flow simulations using the volume of fluid method. A pore network can be employed to quickly identify the differences caused by different fiber curvatures and to determine the necessity for further observation.

To obtain the pore networks of the GDLs in this work, two open-source tools, PoreSpy \cite{gostick2019porespy} and OpenPNM \cite{gostick2016openpnm} are adopted. GDL structures are transferred to binarized volume voxels, labeled with 0 (pore) and 1 (solid). Each cube voxel size uses a resolution of 1E-6 m. Followed by a granulometric analysis using \textit{local\_thickness} function in PoreSpy and a watershed segmentation using a sub-network of an over-segmented watershed algorithm \cite{gostick2017versatile}, the pore region, delineated by watershed segmentation, is connected to form the pore network. The watershed is represented by throats, serving to denote the proximity between two adjoining pore cells. The pores and throats in actual GDL structures are arbitrary shapes. To enhance clarity, pores are illustrated as spherical balls, and throats are depicted as cylinders. The maximal inscribed diameter represents the pore and throat diameter values. Coordination number is introduced in a pore network to describe the number of neighbors of each pore, in other words, the larger value means more connected pores.  Additional significant characteristics of the network, such as pore volume, centroid coordinates, inscribed sphere diameter, throat area, perimeter, centroid, and length, are beyond the scope of this discussion but are comprehensively introduced in a previous study \cite{gostick2017versatile}.

\section{Results and discussion}

\subsection{Mesh sensitivity analysis}
 
\begin{figure}[H]
    \centering
    \subfigure[]{\includegraphics[width=0.4\textwidth]{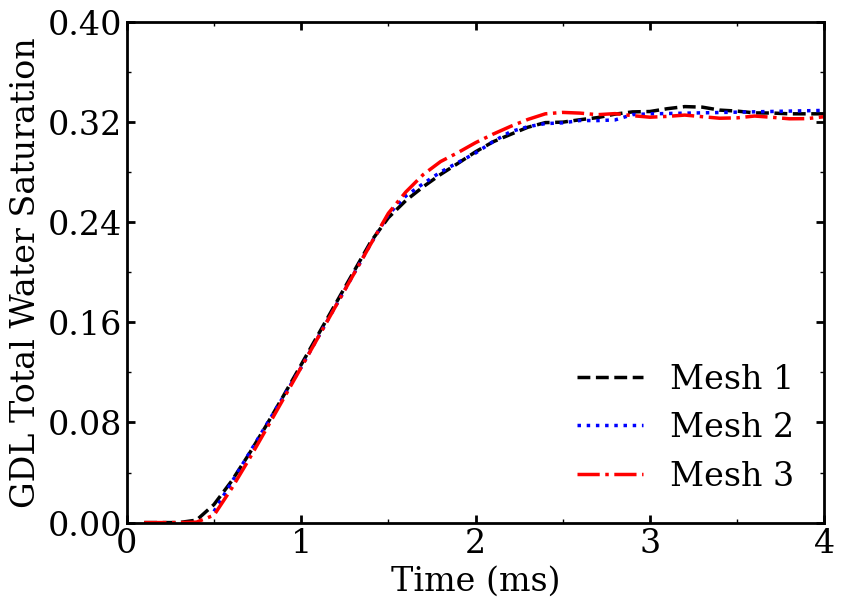}}
    \subfigure[]{\includegraphics[width=0.4\textwidth]{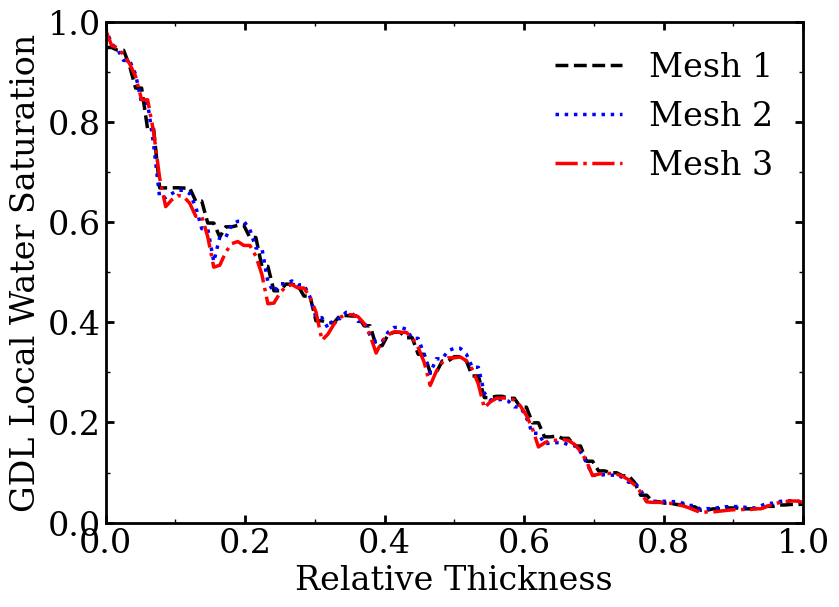}}
    \subfigure[]{\includegraphics[width=0.8\textwidth]{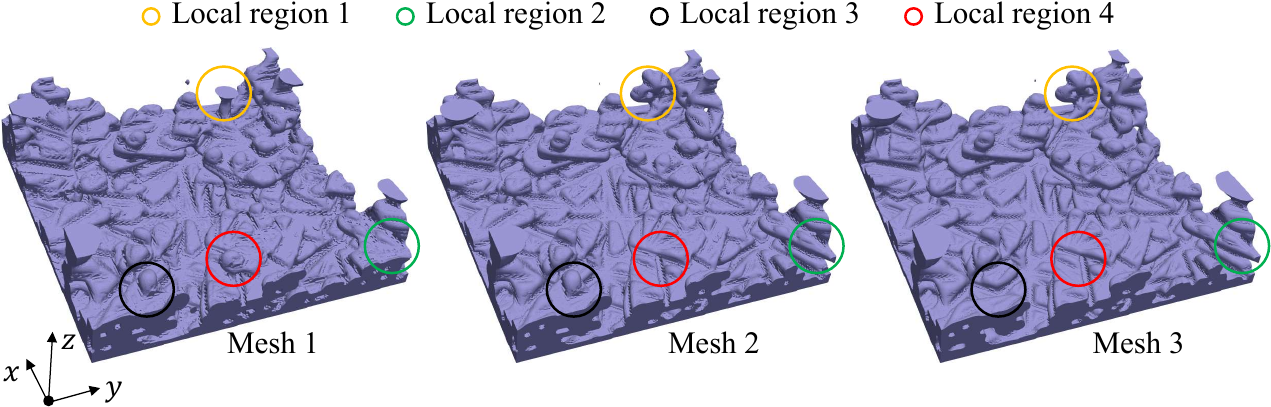}}
    \caption{Mesh sensitivity results, (a) Time-dependent total water saturation in GDL. (b) Through-plane local water saturation at $t=4$ ms. (c) Water distribution in the GDL at $t=4$ ms.}
    \label{MeshSensitivity}
\end{figure}

The mesh was generated using the SnappyHexMesh tool in OpenFOAM. The entire grid is dominated by hexahedral grids. As shown in Table \ref{meshtable}, three different mesh resolutions are used for mesh independence analysis. Mesh 1 is regarded as base mesh, and Mesh 2 and Mesh 3 are sequentially refined with a ratio of 1.3 in each dimension of GC and GD. Compared with those mesh grid resolutions in previous studies \cite{niblett2020two, andersson2018interface, jiao2020water}, it can be seen that the designed coarse grid (Mesh 1) in the GC and GDL in the present study is still comparable to the results in previous studies. Figure \ref{MeshSensitivity}(a-b) shows the total water saturation and planar water saturation of GDL over time. The three grids show comparable results. However, according to the water distribution in Fig. \ref{MeshSensitivity}(c). The amount of water accumulated in the upper two corners of Mesh 1 is significantly different. An additional flow path appears near the right corner compared to the other two finer meshes, which should result from the coarse mesh. Considering the mesh accuracy and computation load, the grid resolution in Mesh 2 was used in subsequent studies. It should be mentioned that in this mesh-independent study, the GDL water fraction initialization was based on a bottom surface rather than on the entire thin region in the GDL bottom in the later studies. This is the reason that water takes some time to start accumulating in the GDL region in Fig. \ref{MeshSensitivity}(a). 
\begin{table}[H]
\small
    \centering
    \caption{Mesh comparison with different grid resolution.}
    \begin{tabular}{llll}
        \hline
         &  Mesh cells& Mesh resolution in GC & Mesh resolution in GDL\\ \hline
       Mesh 1  &3.4 Million&  10.87 $\mu m$ $\times$ 10.87 $\mu m$ $\times$ 10.87 $\mu m$ & 2.72 $\mu m$ $\times$ 2.72 $\mu m$ $\times$ 1.72 $\mu m$ \\ 
       Mesh 2  &7.2 Million& 8.3 $\mu m$ $\times$ 8.3 $\mu m$ $\times$ 8.3 $\mu m$ & 2.08 $\mu m$ $\times$ 2.08 $\mu m$ $\times$ 1.46 $\mu m$ \\  
    Mesh 3 &15 Million&  6.4 $\mu m$ $\times$ 6.4 $\mu m$ $\times$ 6.4 $\mu m$ & 1.6 $\mu m$ $\times$ 1.6 $\mu m$ $\times$ 1.05 $\mu m$ \\
Ref. \cite{niblett2020two} &-& $\approx$ 13 $\mu m$ $\times$ 13 $\mu m$ $\times$ 13 $\mu m$ & $\approx$ 6.8 $\mu m$ $\times$ 6.8 $\mu m$ $\times$ 6.8 $\mu m$ \\
Ref. \cite{andersson2018interface} &-&  $\approx$ 25 $\mu m$ $\times$ 25 $\mu m$ $\times$ 25 $\mu m$ & - \\ 
Ref. \cite{jiao2020water} &-&  - & $\approx$ 3 $\mu m$ $\times$ 3 $\mu m$ $\times$ 1.47 $\mu m$ \\ 
    \hline
    \end{tabular}
    \label{meshtable}
\end{table}

\subsection{Model validation}
\begin{figure}[H]
    \centering
    \subfigure[A rising bubble case]{\includegraphics[width = 0.45\textwidth]{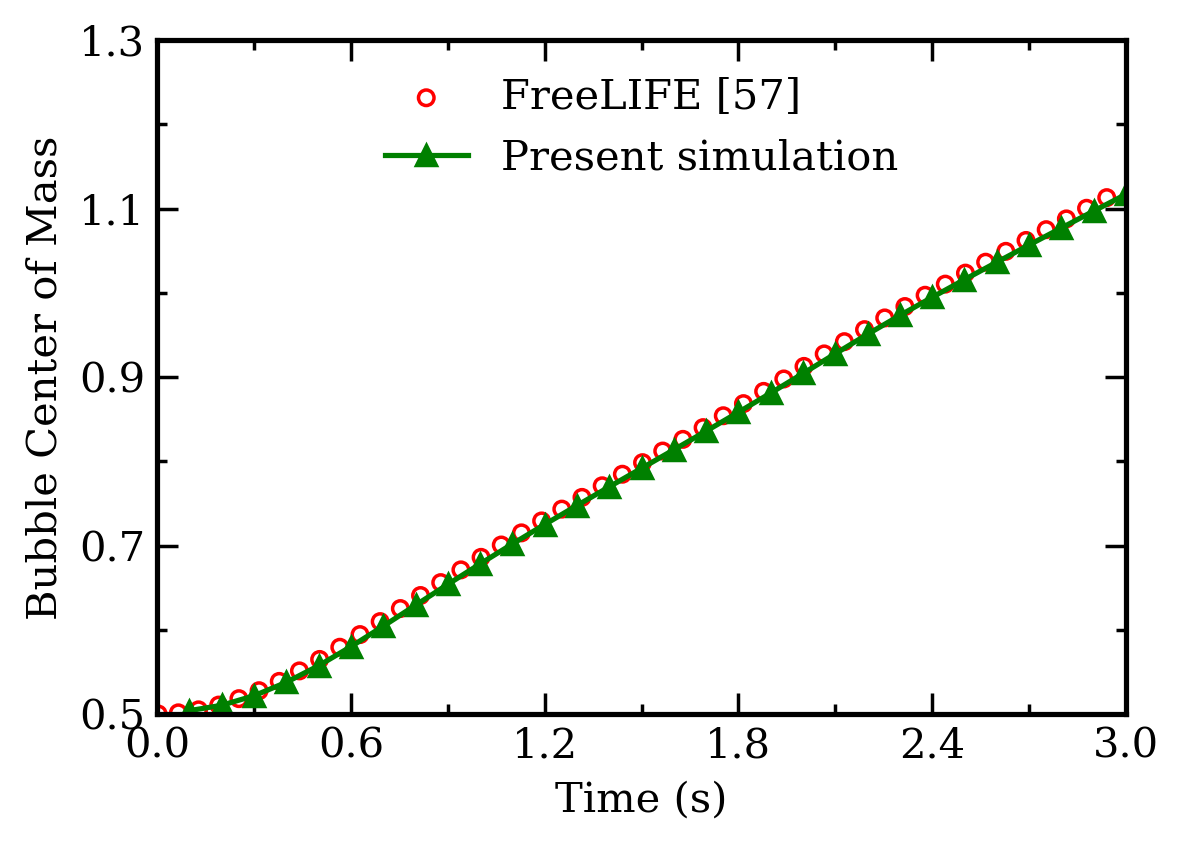}}
    \subfigure[A GDL-GC system case]{\includegraphics[width = 0.46\textwidth]{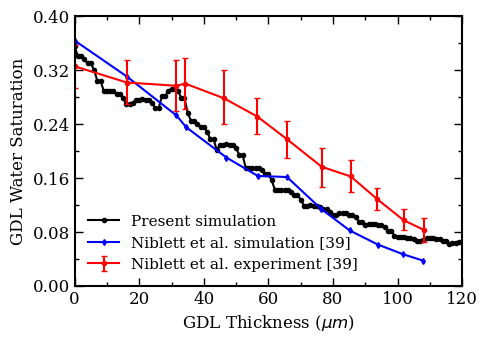}}
    \caption{(a) Comparison of the center of mass of a single rising bubble simulation between the fluid volume method in OpenFOAM and the level set-based simulation method FreeLIFE \cite{hysing2007proposal}. (b) In a GDL-GC system, comparison of the GDL through-plane water local saturation in the present work with both experimental and numerical results in \cite{niblett2020two}. The simulation results in both studies are all extracted from the same simulation time step, $t = 4.5$ $ms$.}
    \label{ModVal}
\end{figure}

\begin{figure}[H]
   \subfigure[]{\includegraphics[width = 0.23\textwidth]{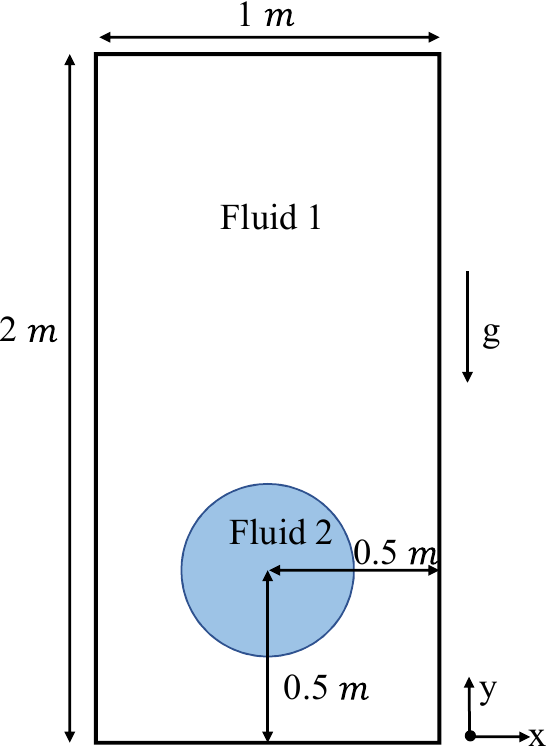}}
  \subfigure[]{\includegraphics[width = 0.3\textwidth]{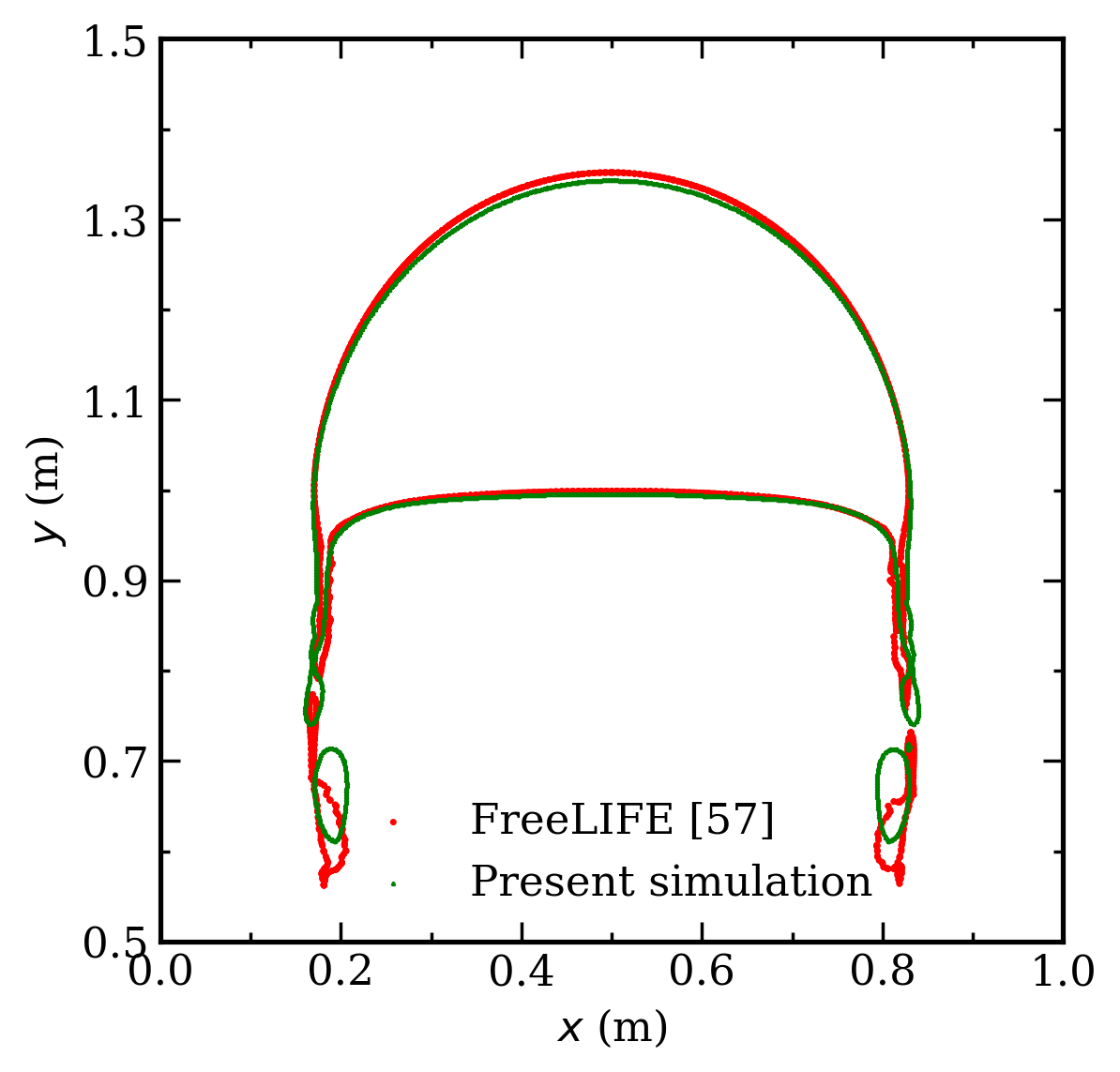}}
  \subfigure[]{\includegraphics[width = 0.43\textwidth]{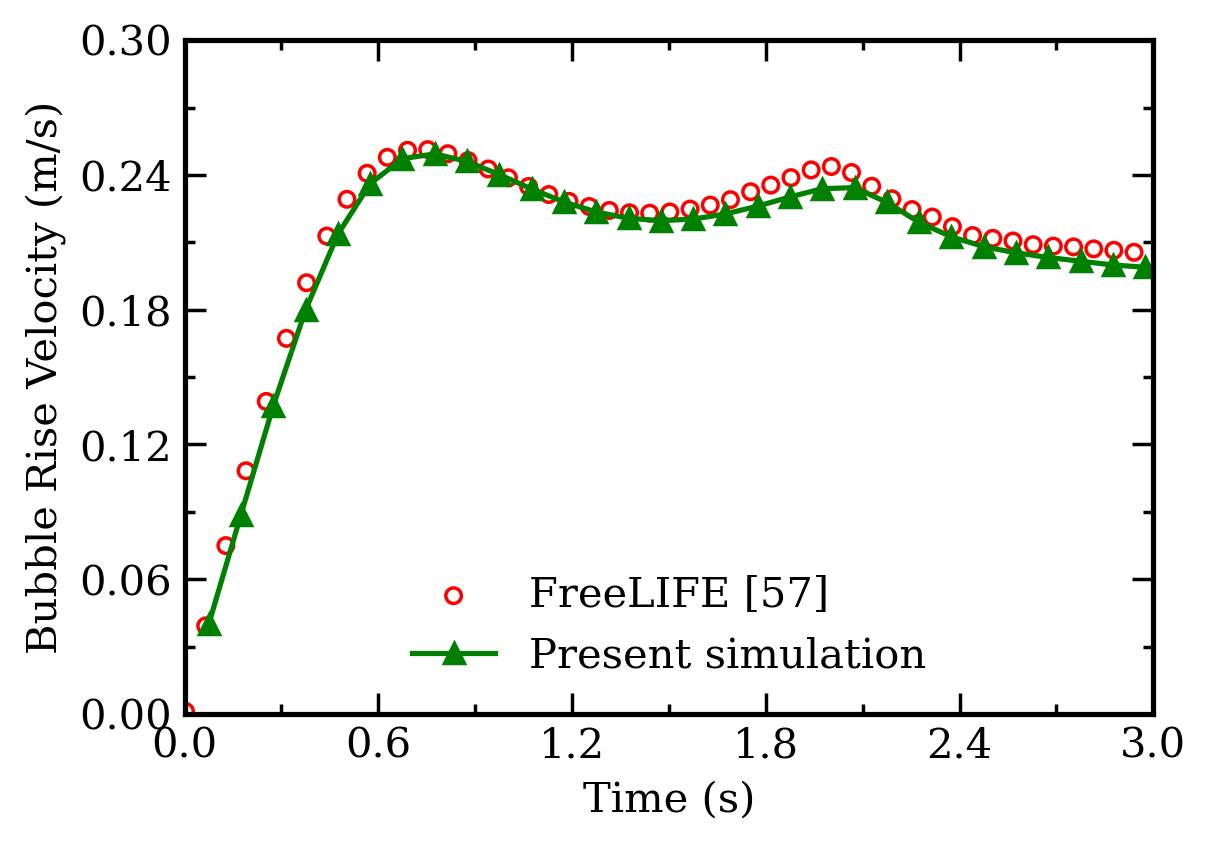}}
   \subfigure[]{\includegraphics[width = 1\textwidth]{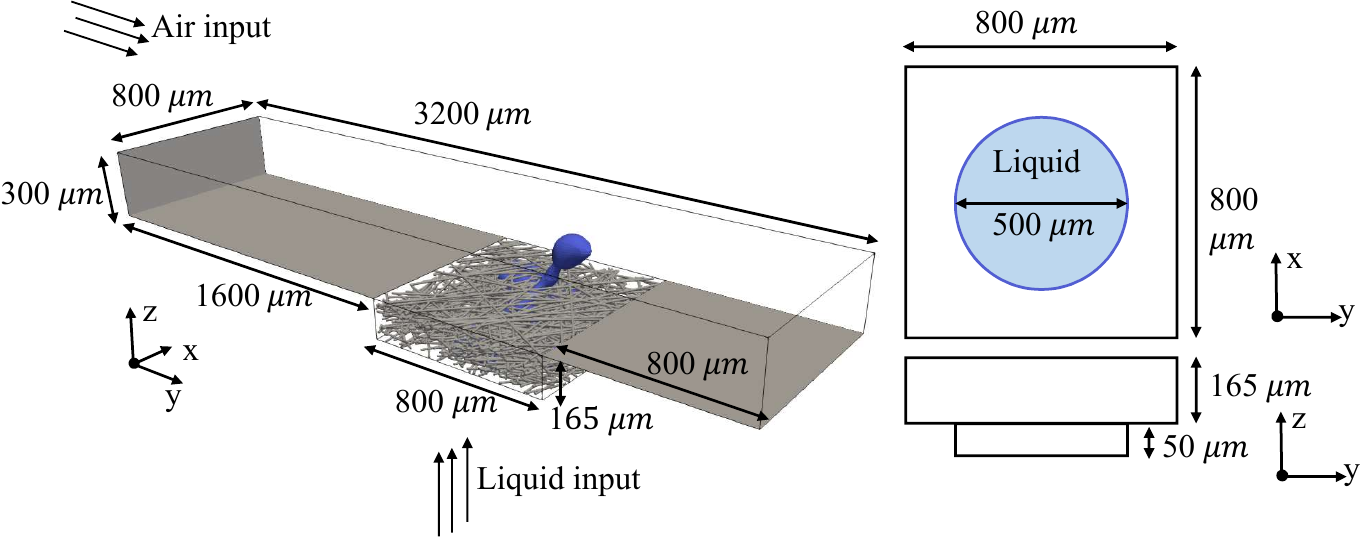}}
    \caption{(a) The geometry for simulation model comparison on a rising bubble. (b-c) Comparison of bubble shape and bubble rise velocity between two simulation methods. (d) The geometry and dimensions for validation simulation in a GDL and GC system, including a long rectangular GC and a smaller rectangular  GDL in the middle as well as a cylindrical water inlet in the bottom. The right two are the top view and side view of the central part with GDL.}
    \label{Validations}
\end{figure}

The volume of fluid method implemented in OpenFOAM has been validated in different situations, for example, bubble rise \cite{gamet2020validation, ghanavati2023numerical}, dam break \cite{vosoughi2020experimental, meng2022three}, multi-scale channel flow \cite{andersson2018modeling, malekzadeh2015investigation}. In this study, a two-dimensional rising bubble is simulated with the same solver schemes as those utilized in this study. The results are compared with that of another simulation method FreeLIFE \cite{hysing2007proposal}, developed based on the level-set method. Both simulations exhibit satisfactory similarity concerning the bubble center of mass. Detailed simulation geometry and two more comparisons of bubble shape and bubble rise velocity are presented in Fig. \ref{Validations}(a-c). Besides, to validate the application of the volume of fluid method in the GDL and GC two-phase flow simulation, an initial comparison between experimental and numerical results has shown a similar variation trend in the local water saturation variation \cite{yang2023numerical}. The GC water behavior has also been compared with synchrotron-based X-ray radiography and tomography imaging \cite{andersson2018modeling}. For such a complex GDL microstructure, an ideal comparison should keep the same GDL topology in both experiment and simulation. Both experimental image-based porous structure reconstructions and water saturation measurements are mostly based on high-resolution scanning techniques and good image processing approaches, which increases the difficulty of giving good precise results. Niblett et al. \cite{niblett2020two} have used the same GDL geometry for both experimental and numerical studies, as exhibited in Fig. \ref{ModVal}(b). Both methods show a similar GDL through-plane water saturation trend, starting to decline from around 0.32. However, it can be seen that there is still some difference between the experiment and simulation despite keeping the same geometry. In the present research, a comparable analysis has also been carried out based on a stochastic straight-fiber GDL reconstruction by controlling as many parameters as possible same as the experimental setup in \cite{niblett2020two}. The simulation geometry is shown in Fig. \ref{Validations}(d), and vital parameters have been listed in Table \ref{Modvalmy_label} in \ref{Appendix II}. The present simulation also shows comparable results to the previous results, which further illustrates the viability of the simulation method.

\subsection{GDL structure comparison}
\begin{figure}[H]
    \centering
    {\includegraphics[width = 0.9\textwidth]{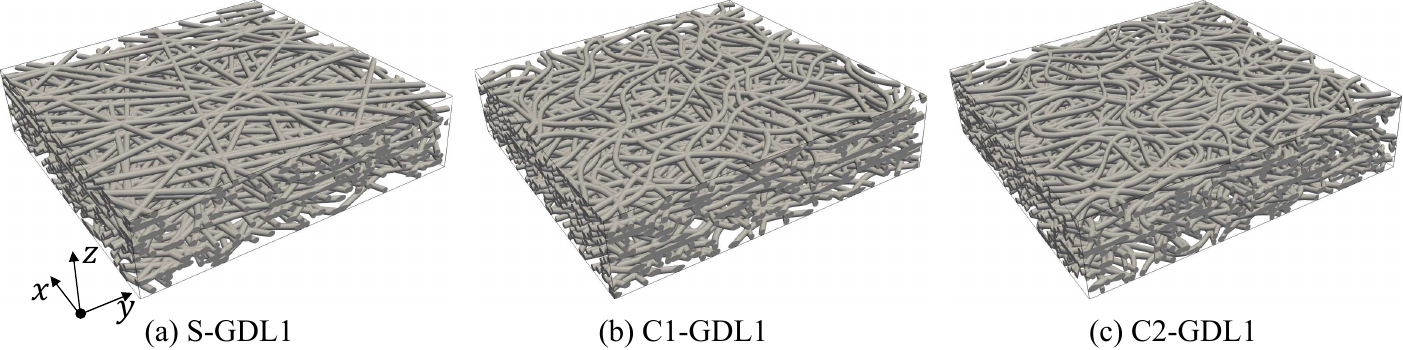}}
    \caption{Three representative GDL configurations with different fiber curvature. (a) S-GDL1 (straight fibers with $b=0, f_r = 0$); (b) C1-GDL1 (curved fibers with $b=0.2, f_r = 2.5$); (c) C2-GDL1 (curved fibers with $b=0.4, f_r = 2.5$).}
    \label{GDLSamples}
\end{figure}

Given the inherent uncontrollable randomness observed in the GDL during manufacturing, such as variations in fiber orientation, achieving a completely ordered structure relying on current techniques is still challenging \cite{liu2023numerical}. Consequently, this investigation maintains the random fiber orientation instead of pursuing a theoretically ordered configuration. Three distinct fiber curvatures, characterized by the parameters $b=0, f_r=0$; $b=0.2, f_r=2.5$; and $b=0.4, f_r=2.5$, have been employed to reconstruct three types of GDLs. To mitigate potential misinterpretations arising from this randomness, each type of GDL reconstruction comprises four samples, which are generated using a consistent procedure. Besides, these GDLs undergo reconstruction through a layer-by-layer stacking approach. Table \ref{Different curvature} provides data on bulk porosity and five layer porosity values for both physical and virtual GDLs. Layer 1 is situated at the GDL bottom and Layer 13 is positioned near the GDL/GC interface. Note that the capital letters S, C1, and C2 represent GDL fibers with straight, curved, and more curved shapes, respectively. The maximum deviation of bulk porosity values from those of the authentic GDL H2315 is below 1\%, while the deviation of layer porosity is under 2\%. These discrepancies fall within acceptable limits for this investigation. Achieving a smaller tolerance error requires a more strict constraint at the expense of an extended generation time. Additionally, the observed similarity in layer and bulk porosity illustrates the controllability of the layer-by-layer stacking strategy, which shows the potential for future fabrication of such structures. Representative illustrations of the first sample for each type are presented in Fig. \ref{GDLSamples}(a-c). From S-GDL1 to C2-GDL1, an increasing presence of curved fibers is observable.

\begin{table}[H]
\footnotesize
    \centering
     \caption{Structural bulk and layer porosity in GDLs with varied fiber curvatures (S, C1, and C2).}
    \begin{tabular}{p{1.9cm}p{2.7cm}p{1.3cm}p{1.3cm}p{1.3cm}p{1.3cm}p{1.3cm}p{1.3cm}}
    \hline
         GDL name&  Fiber curvature& Bulk porosity & Layer 1 porosity  & Layer 4 porosity  & Layer 7 porosity& Layer 10 porosity& Layer 13 porosity\\ \hline
         H2315 GDL& - & 0.6985&0.7165&0.6897&0.7075&0.6688&0.7350\\
         S-GDL1& $b=0,f_r = 0$& 0.7039&0.7282&0.6967&0.7100&0.6780&0.7404 \\
         S-GDL2& & 0.7012&0.7189&0.6908&0.7080&0.6757&0.7418\\
         S-GDL3& & 0.7027&0.7264&0.6870&0.7107&0.6759&0.7435\\
         S-GDL4& & 0.7019&0.7246&0.6911&0.7116&0.6769&0.7382\\
         C1-GDL1&$b=0.2,f_r = 2.5$& 0.7006&0.7268&0.6915&0.7049&0.6691&0.7366\\
         C1-GDL2& & 0.7010&0.7168&0.6957&0.7138&0.6729&0.7395\\
         C1-GDL3& & 0.7021&0.7268&0.6914&0.7099&0.6686&0.7444\\
         C1-GDL4& & 0.7021&0.7263&0.6909&0.7166&0.6703&0.7436\\
         C2-GDL1&$b=0.4,f_r = 2.5$ & 0.6990&0.7246&0.6867&0.7081&0.6729&0.7404\\
         C2-GDL2& & 0.6995&0.7148&0.6926&0.7110&0.6701&0.7435\\
         C2-GDL3& &0.6989&0.7253&0.6891&0.7078&0.6685&0.7401\\
         C2-GDL4& & 0.6997&0.7169&0.6866&0.7025&0.6716&0.7364\\\hline
         \multirow{2}{2cm}{\centering Maximum deviation} & Absolute value &0.0055&0.0135&0.0101&0.0075&0.0095&0.0054\\
         &Percentage& 0.78\%&1.85\%&1.45\%&1.05\%&1.40\%&0.70\%\\\hline
    \end{tabular}
    \label{Different curvature}
\end{table}

\subsection{GDL pore networks}\label{porenetwork}

\begin{figure}[H]
    \centering
    {\includegraphics[width = 0.6\textwidth]{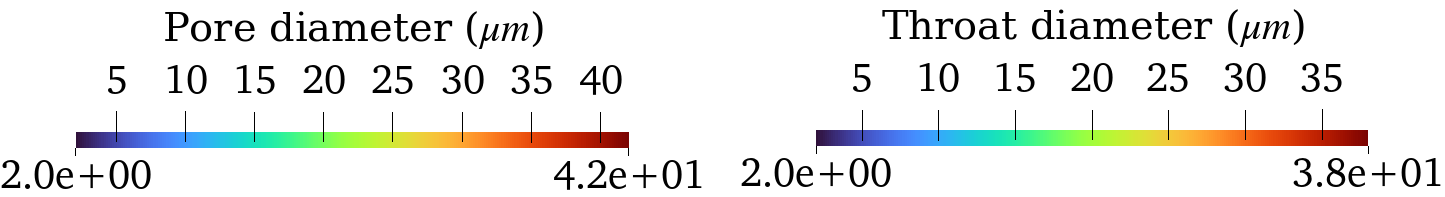}}
    
    {\includegraphics[width = 1\textwidth]{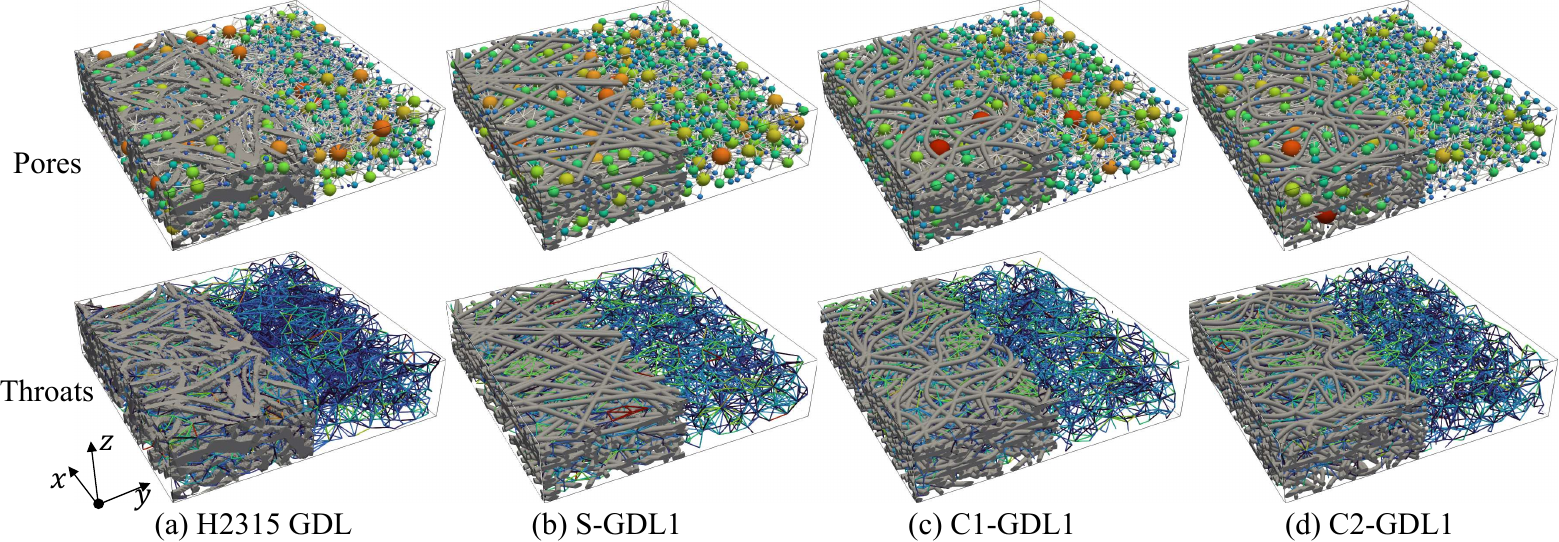}}
    \caption{The spatial pore (the first row) and throat (the second row) distribution in pore networks for H2315 GDL and S-GDL1, C1-GDL1, C2-GDL1, colored by the pore and throat diameter, respectively.}
    \label{GDLTotalPNMPoreDiameter}
\end{figure}

Figure \ref{GDLTotalPNMPoreDiameter} displays the spatial distribution of pores and throats in the pore networks of H2315 GDL, S-GDL1, C1-GDL1, and C2-GDL1, serving as representative samples in each type. The H2315 GDL exhibits denser pores and throats compared to the other three numerically regenerated GDLs, evident in the prevalence of red-colored pores and denser blue-colored throats in Fig. \ref{GDLTotalPNMPoreDiameter}(a). Note that the displayed pore network is fully interconnected, excluding isolated individual pores and locally connected pores. Thus, dense throats indicate robust connectivity among pores. Enlarging GDL fiber curvature in Fig. \ref{GDLTotalPNMPoreDiameter}(b-d), an increasing trend in small pores and throat density, as well as a reduction of large pores can be observed. 

\begin{figure}[H]
    \centering
    \subfigure[H2315 GDL]{\includegraphics[width = 0.8\textwidth]{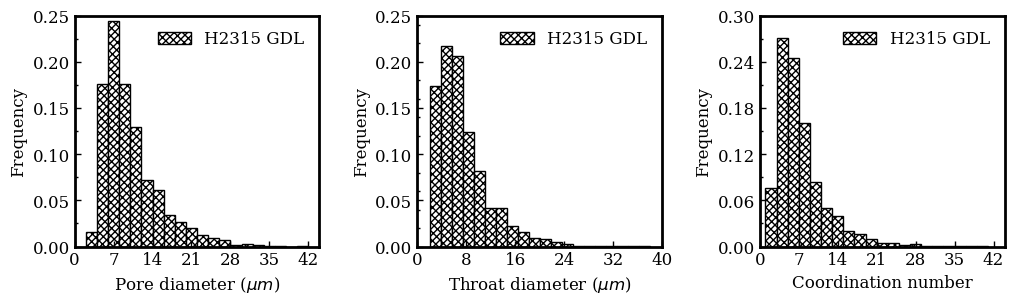}}
    \subfigure[S-GDLs]{\includegraphics[width = 0.8\textwidth]{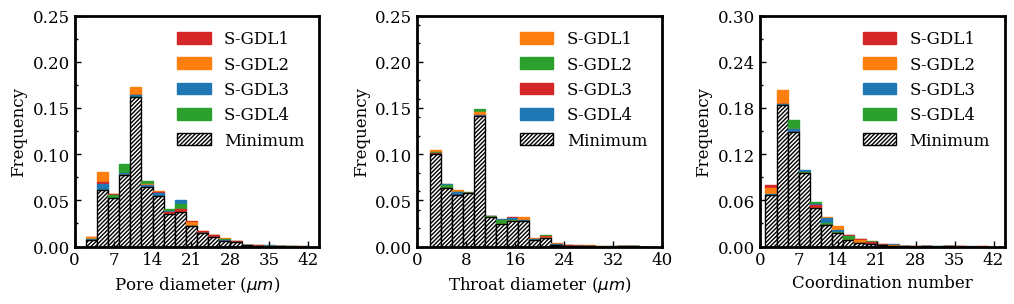}}
    \subfigure[C1-GDLs]{\includegraphics[width = 0.8\textwidth]{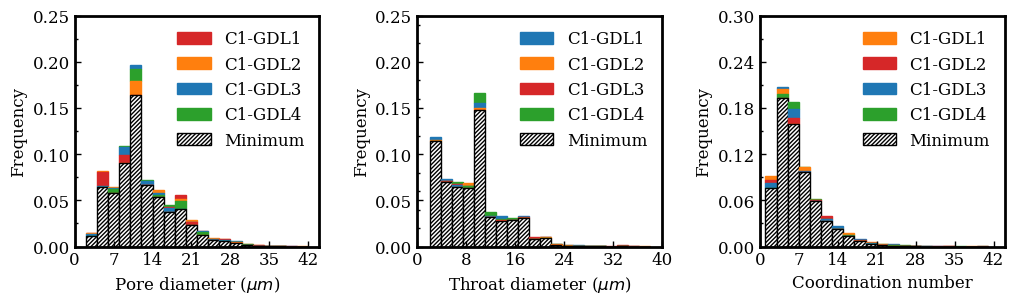}}
    \subfigure[C2-GDLs]{\includegraphics[width = 0.8\textwidth]{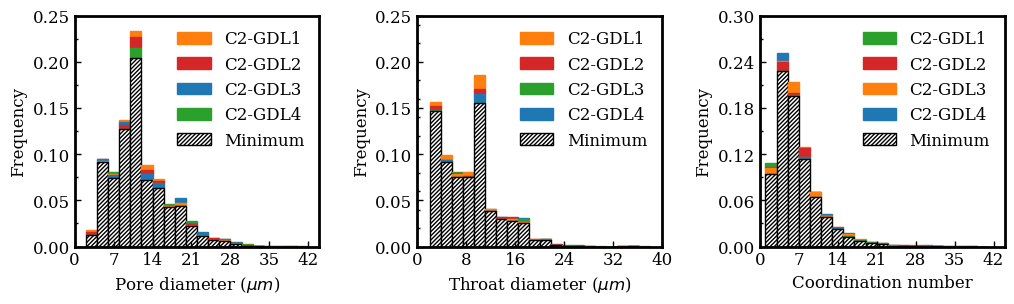}}
    \caption{Frequency distribution of pore diameter, throat diameter, and coordination number for pore networks of experimental and all reconstructed GDL. Their varying ranges are respectively [2-42 $\mu m$], [2-38 $\mu m$], and [1-41]. The legend 'Minimum' in Fig. \ref{GDLTotalPNMFrequency}(b-d) represents the minimal value in each subregion of the four samples in each type of GDL. The frequency of each case equals the sum of its color region and all below regions.}
    \label{GDLTotalPNMFrequency}
\end{figure}

To extract detailed difference among these pore networks, Fig. \ref{GDLTotalPNMFrequency}(a-d) further distinguishes the frequency distribution of pores, throats, and an additional parameter, coordination number. Thirty uniform sub-regions are utilized for frequency statistics of each parameter. A higher sub-region division number yields a comparable distribution, indicating the independence of subregion numbers in the analysis. It is noteworthy that the specific frequencies for pore diameter, throat diameter, and coordination number in various scenarios are derived using different base constants, namely 3600, 13000, and 3600, respectively. In Fig. \ref{GDLTotalPNMFrequency}(b-d), the term 'Minimum' denotes the minimum recorded value within individual subregions across the four distinct samples of each GDL type. Frequency values are determined by summing the corresponding color region and all adjacent subregions below it, ensuring a clear representation of sample frequency without encountering issues of color overlap or transitions. The deviation of each sample from the lowest value within the same category can be calculated. Furthermore, the observed discrepancies in Fig. \ref{GDLTotalPNMFrequency}(b-d) are relatively minor, which suggests that each four stochastically generated GDLs with the same fiber curvature exhibit comparable frequency distributions for pore diameter, throat diameter, and coordination number.

Comparing Fig. \ref{GDLTotalPNMFrequency}(a) and Fig. \ref{GDLTotalPNMFrequency}(b-c), it can be seen that the H2315 GDL sample and all numerically reconstructed GDLs have apparent difference in the three distributions. Notably, the former exhibits a higher frequency peak value and a smaller peak frequency distribution range. In detail, H2315 GDL has the largest pore and throat ratios with pore diameters of 6-8 $\mu m$ and throat diameters of 4-6 $\mu m$. The frequency distribution of pore throat sizes exhibits a sudden rise preceding the peak and then shows a gradual decline thereafter. In comparison, it is found that three types of reconstructed GDLs exhibit larger pore and throat peak frequency distribution regions, and both have a diameter range of 8-10 $\mu m$. Besides, their pore size distribution exhibits a cliff-like decline on both sides of the peak value. Throat size distribution shows the second-largest frequency value in the smallest throat diameter region. The difference between these two peak values seems to decrease with enhancing fiber curvature. The similarities among the three types of GDLs may result from the same layer-by-layer stacking methods. Compared with the pore and throat frequency distribution among all pore networks, the coordination number distribution shows similar variation profiles as the increase in the coordination number, while H2315 GDL almost has the highest frequency value in each region, indicating the best connection features. With the increase of fiber curvature in reconstructed GDLs, the pore connection becomes better. 
Consequently, increasing fiber curvature leads to more pores and throats, thereby better connections.

\subsection{GDL water saturation and capillary pressure}
\begin{figure}[H]
    \centering
    \subfigure[Straight fiber GDLs]{\includegraphics[width = 0.4\textwidth]{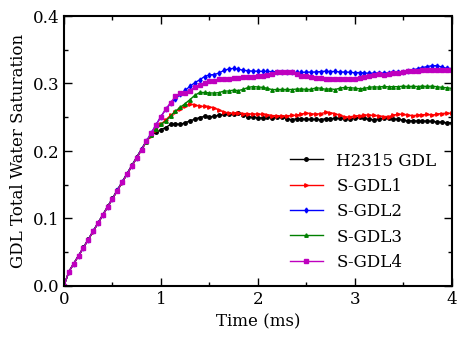}}
    \subfigure[Curved-fiber GDLs, $b=0.2$]{\includegraphics[width = 0.4\textwidth]{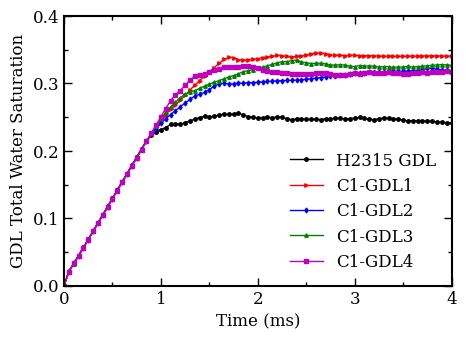}}
    \subfigure[Curved-fiber GDLs, $b=0.4$]{\includegraphics[width = 0.4\textwidth]{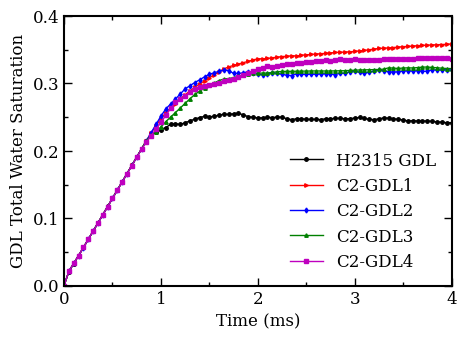}}
    \subfigure[Ensemble average for three types]{\includegraphics[width = 0.4\textwidth]{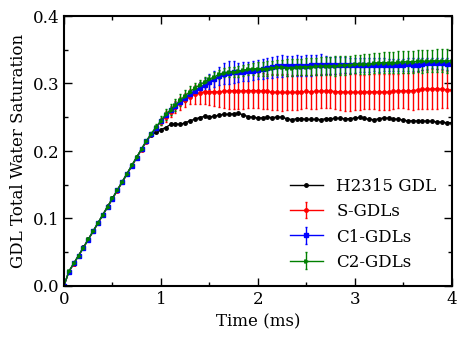}}
    \caption{Time-dependent water saturation state in the GDLs with three distinct fiber curvatures in 4 ms. The simulation results based on a physical GDL H2315 are used for comparison.}
    \label{GDLTotalSa}
\end{figure}

In Fig. \ref{GDLTotalSa}(a-c), the time-dependent total water saturation state is presented for three types of GDLs featuring distinct fiber curvatures, during 4 ms. Prior to 1 ms, all GDLs exhibited comparable rates of water accumulation within the material. However, their water behavior diverges thereafter, despite similarities in fiber diameter, bulk porosity, and through-plane local layer porosity. Table \ref{Breakthrough} compiles the water breakthrough times and corresponding capillary pressures for each GDL. Water breakthrough occurs across all GDLs within the time span of 0.8 ms to 1.1 ms. Notably, there is an approximate upward trend in breakthrough pressure with the increase of fiber curvature. Further specific elaboration on capillary pressure will be provided subsequently. 

Following the breakthrough, the water saturation in GDL H2315 experienced a slight increase before reaching statistical stabilization, which is lower than those within all virtually reconstructed GDLs. Only S-GDL1 shows a similar level of water saturation. The concentrated small pores and throats in H2315 GDL probably contribute to its inside low saturation, as water is forced to flow from large pores. 
In contrast, reconstructed GDLs show a more pronounced increase in water accumulation. This may be due to their larger pore and throat diameters corresponding to the peak frequency. Figure \ref{GDLTotalSa}(a) reveals that S-GDLs achieve statistically stable water saturation within the GDLs after 2 ms. For C1-GDLs, Fig. \ref{GDLTotalSa}(b) shows that C1-GDL3 and C1-GDL4 begin the stabilization around 2.5 ms, while C1-GDL1 and C1-GDL2 continue accumulating water at a slower rate. Four C2-GDLs exhibit gradual water accumulation after 1.5 ms (See Fig. \ref{GDLTotalSa}(c)). Figure \ref{GDLTotalSa}(d) depicts the ensemble average water saturation for different GDL types throughout the specified duration. S-type GDLs exhibit an average water saturation closely resembling that in the physical GDL but show the widest uncertain range. Both C1-type and C2-type GDLs demonstrate comparable average values and exceed the maximum of the S-GDLs. Consequently, a noticeable distinction emerges between straight-fiber and curved-fiber GDLs. Additionally, a subtle divergence in water behavior is apparent between C1-type and C2-type GDLs. with increasing the fiber curvature, a slow increase in total water saturation after 2 ms becomes more general, which may related to the increased small pores and throats.

\begin{table}[H]
\footnotesize
    \centering
     \caption{GDL breakthrough time and corresponding capillary pressure}
    \begin{tabular}{p{1.9cm}p{2.7cm}p{2.3cm}}
    \hline
         GDL name&  Breakthrough time (ms)& Breakthrough pressure (kPa) \\ \hline
         H2315 GDL& 0.85 &5.26 \\
         S-GDL1& 0.85&  5.83\\
         S-GDL2& 1.05& 6.06\\
         S-GDL3& 0.85& 5.58\\
         S-GDL4& 1& 6.05\\
         C1-GDL1&0.85& 5.94\\
         C1-GDL2& 0.9& 6.25\\
         C1-GDL3& 1.1& 6.37\\
         C1-GDL4& 1.1& 6.58\\
         C2-GDL1&0.85& 6.72\\
         C2-GDL2& 1.05& 6.91\\
         C2-GDL3& 0.85&6.51\\
         C2-GDL4& 0.8& 7.08\\\hline
    \end{tabular}
    \label{Breakthrough}
\end{table}

The discrepancy among four samples of each type of GDL can be attributed to the irregularity in the distribution and connection of various sizes of pores, as we have discussed in Section \ref{porenetwork}, which is challenging to regulate in stochastic generation and even actual manufacturing. This observation also highlights the challenge of drawing definitive conclusions in two-phase flow studies solely based on one randomly reconstructed GDL. Therefore, the ensemble average water saturation of three kinds of GDLs is shown in Fig. \ref{GDLTotalSa}(d). It should be mentioned that the same type of physical GDL samples also should have a difference in inside water saturation. Due to the structure limitation, this difference is not shown for the physical GDL in this study. The error bar plot indicates that the straight fiber GDLs bring a larger deviation compared with the curved fiber GDLs even if it provides the possibility to be close to the water dynamics within the real GDL. On the contrary, the curved-fiber GDL samples give higher ensemble average water saturation, but the deviation is relatively smaller, which decreases the uncertainty.

\begin{figure}[H]
    \centering
    \subfigure[Straight fiber GDLs]{\includegraphics[width = 0.4\textwidth]{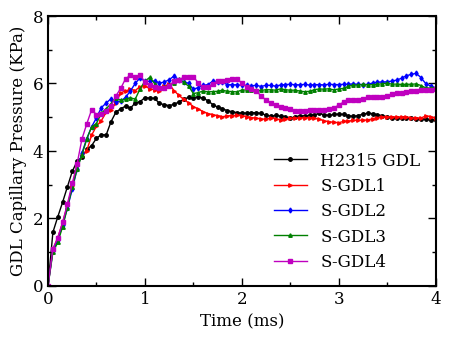}}
    \subfigure[Curved-fiber GDLs, $b=0.2$]{\includegraphics[width = 0.4\textwidth]{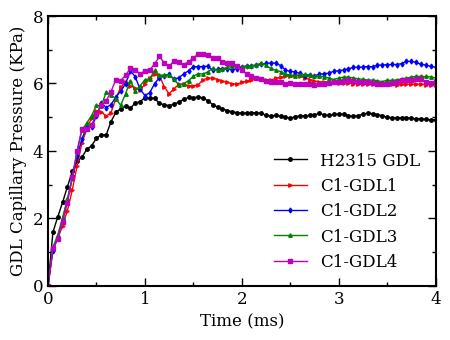}}
    \subfigure[Curved-fiber GDLs, $b=0.4$]{\includegraphics[width = 0.4\textwidth]{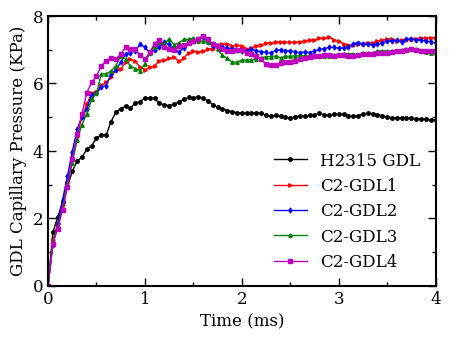}}
    \subfigure[Ensemble average for three types]{\includegraphics[width = 0.4\textwidth]{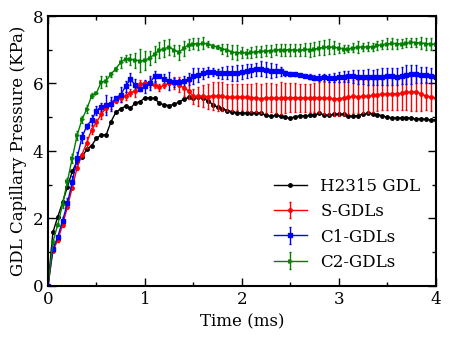}}
    \caption{Time-dependent capillary pressure in the GDLs with three distinct fiber curvatures in 4 ms. The simulation results based on a physical GDL H2315 are used for comparison.}
    \label{GDLTotalCa}
\end{figure}

Figure \ref{GDLTotalCa} displays the time-varying capillary pressure in different GDLs. Before the water breakthrough, the GDL sample showed a similar quick increase as the other three samples of the same type. All of them begin to have a stabilization trend from $t = 1$ ms. However, different levels of capillary pressure fluctuations can be seen, rather than the relatively smooth changes in water saturation. Since the breakthrough, different levels of pressure overshoot as well as relatively rapid fluctuation during 0.8 ms and 1.5 ms also can be observed. Therefore, the capillary pressure is more sensitive to the pore structure difference and flow evolution compared with the total water saturation. 

Similar to the total water saturation, the four straight-fiber GDLs have a great distinction among them, while the S-GDL1 has a close capillary pressure variation to the experimental GDL H2315. Whereas, the other GDL samples have higher capillary pressure than the H2315 GDL. As the fiber curvature increases, from S-type to C2-type, the GDL capillary pressure has a visibly increasing trend compared with the trend of total water saturation, see Fig. \ref{GDLTotalCa}(a-c). However, a slightly higher capillary pressure can be found at the beginning period of water invasion (before around 0.4 ms) in Fig. \ref{GDLTotalCa}(a-b) and becomes very little in Fig. \ref{GDLTotalCa}(c). Figure \ref{GDLTotalCa}(d) shows the ensemble average capillary pressure variation over time. Straight-fiber GDLs have the biggest error and lowest value among the three types. Both curved-fiber GDLs have smaller errors, while the C2-GDLs have the largest capillary pressure, which has an apparent difference from that of the C1-GDLs, unlike the ensemble average total water saturation in Fig. \ref{GDLTotalCa}(d). Therefore, the close relationship between capillary pressure and total water saturation is embodied.

With the above phenomenon, an extended consideration is that even though the whole in-plane local porosity of an actual GDL can be kept relatively uniform, the uniformity of the inside water distribution is still hard to guarantee. Non-uniform water distribution will lead to non-uniform oxygen distribution, thereby non-uniform current density distribution. To improve the controllability of GDL species transport properties, one possible solution is to try to control the spatial distribution of pores considering both size and connection. Even if the curved-fiber GDL water behavior is far away from the real one in the present work, the smaller water saturation deviation among different samples also shows potential in future GDL optimal designs.

Combining the results of Fig. \ref{GDLTotalPNMFrequency} and Fig. \ref{GDLTotalCa}, a positive correlation is observed between the pore network and capillary pressure in reconstructed GDLs. Higher fiber curvature in these GDLs results in an increasing frequency of small pores and throats, contributing to elevated capillary pressure. H2315 GDL exhibits dissimilar characteristics compared to C2-GDLs, despite both having high frequencies in smaller pore and throat regions. This discrepancy can be attributed to the higher frequency of coordination numbers in H2315 GDL, indicating improved pore connections. Consequently, there is an enhanced likelihood of larger pores establishing connections with efficient water pathways, mitigating the influence of smaller pores. Additionally, variations in water saturation and capillary pressure between H2315 GDL and C2-type GDL result from dominant pore diameter regions, despite relatively minor differences in coordination number distribution. Thus, the importance of water transport is not solely dependent on the number of pores and throats; the coordination number, reflective of GDL connection quality, also holds significance.

\begin{figure}[H]
    \centering
    {\includegraphics[width = 1\textwidth]{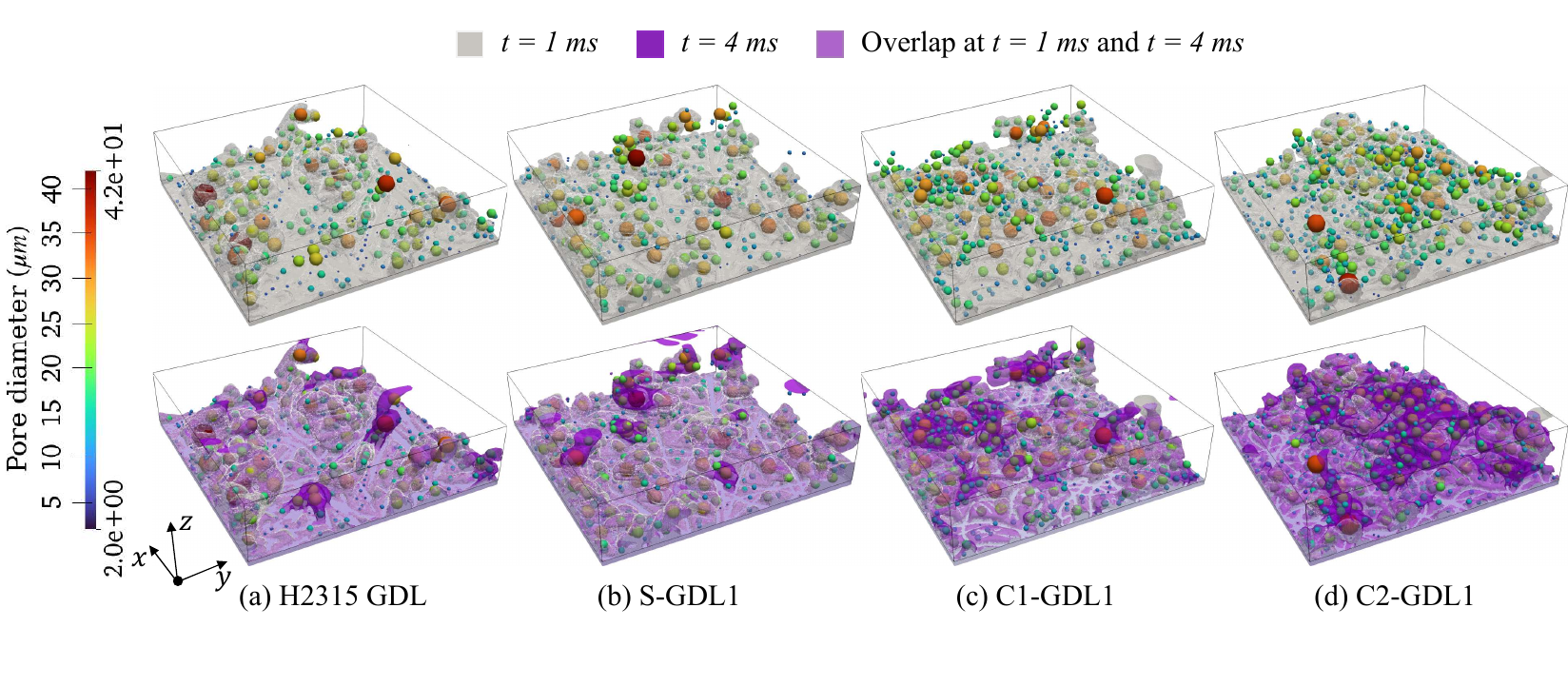}}
    \caption{Combination of filtered pores and GDL water distribution state at 1 ms (gray surface) and 4 ms (purple surface). The gray color, dark purple color, and light purple color represent the water saturation at 1 ms, the water saturation at 4 ms, and the overlap water distribution at two timesteps, respectively. The pores are colored with the pore diameter size (see left color bar).}
    \label{GDLTotalSaDifferenceWithPores}
\end{figure}

Figure \ref{GDLTotalSaDifferenceWithPores} is used to further investigate the relationship between saturated water in Fig. \ref{GDLTotalSa} and the pore network. The water saturation at 1 ms and 4 ms is selected to represent before and after water breakthrough and colored in gray and purple, respectively. The pore networks in Fig. \ref{GDLTotalPNMPoreDiameter} are filtered by removing the pores outside the water accumulation topology at 4 ms. The top four panels in Fig. \ref{GDLTotalSaDifferenceWithPores} represent the combination of filtered pores and water saturation (gray) at 1 ms. It can be seen that most of the pores are included in the gray topology while a few big pores (in H2315-GDL and S-GDL1) and small-pore clusters (in C1-GDL1 and C2-GDL2) are outside. It can be seen that water quickly occupies the regions clustered by large pores, which results in non-uniform spatial water distribution across GDLs. The subregion without water rising up is found with very small pores. In addition, with the increase of fiber curvature, an increasing number of small pores (for example, blue color pores) are included in the water topology. Even though the water saturation at this moment is similar among the four samples. However, the small pores are found to contribute to the increasing capillary pressure. The bottom four figures show the combination of filtered pores and the water saturation over two time. The transitional light purple color represents the overlap in water saturation between these two moments, indicating smaller water changes in these regions. Dark purple represents the new water saturation. It is found that delayed water saturation after 1 ms in H2315 GDL and S-GDL1 is primarily attributed to both the emergence of a new breakthrough flow path connected by big pores and water accumulation in big pore clusters, see Fig. \ref{GDLTotalSaDifferenceWithPores}(a-b). According to Fig. \ref{GDLTotalSaDifferenceWithPores}(c-d), increasing fiber curvature in C1-GDL1 and C2-GDL1 correlates with noticeable water accumulation in middle-size pores. The delayed water accumulation in unique larger pores may result from difficult water breakthrough within prior connected small pores and throats.

\begin{figure}[H]
    \centering
   \includegraphics[width = 0.95\textwidth]{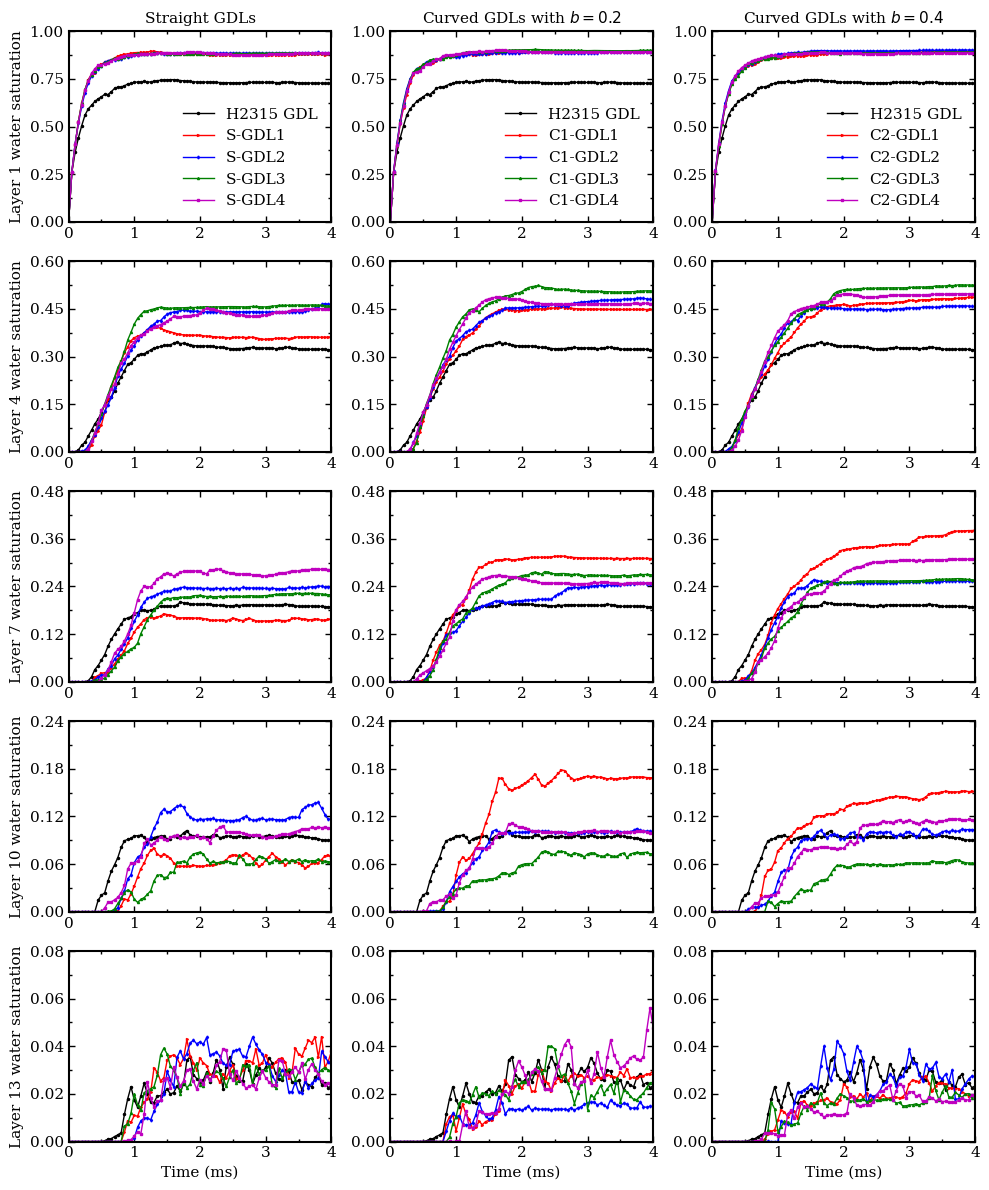}
    \caption{Layer water saturation GDLs with different fiber curvature. Layer 1 is close to the GDL bottom and Layer 13 is near the GDL/GC interface.}
    \label{TotalLayerSa}
\end{figure}
 
Furthermore, Fig. \ref{TotalLayerSa} displays the local layer water saturation at five layers (1, 4, 7, 10, 13). From the GDL bottom (Layer 1) to the GDL top (Layer 13), the local layer water saturation gradually decreases in all cases. Even though the similar through-plane layer porosity among all GDLs, as displayed in Table \ref{Different curvature}, the individual GDL in each type gradually shows a larger difference from the other samples over time, apart from the similar variations in layer 1. A type of GDL is usually designed and manufactured with distinct through-plane porosity distributions. For example, the Toray TGP-H type \cite{fishman2011heterogeneous} shows higher porosity on both sides and lower porosity in the middle. Besides, some people also try to design gradient porosity distribution \cite{sim2022effects}. However, we find that all the GDLs here show different through-plane layer water variations. In each type, a GDL exhibiting reduced water saturation in a specific lower layer compared to other specimens does not necessarily maintain lower water saturation in subsequent layers relative to other samples. For instance, S-GDL3 displays the least water saturation in Layer 4 among the four samples, yet exhibits the highest water saturation in Layer 7, with middle water saturation level observed in Layer 10. This finding means the layer water distribution can not be well controlled by a specific through-plane porosity distribution. It can be seen that different sizes of pores are randomly located in different layers in Fig. \ref{GDLTotalPNMPoreDiameter}, which is also a key factor for water distribution. Not only the layer porosity but also a possible control of the fiber location and orientation which can contribute to the layer pore size adjustment and through-plane pore connectivity should be paid attention to in future work. 

In addition, for the GDLs with curvature $b = 0.4$, a slight increase in total water saturation after breakthrough in Fig. \ref{GDLTotalSa}(c) is found resulting from the increase in the 7th and 10th layers. Starting from Layer 7, H2315 GDL has a faster water accumulation rate than that of all other reconstructed GDLs before water breakthrough, followed by quick stabilization. This may result from its good connections among big pores and short breakthrough pathways. As the increase of fiber curvature, an increasing trend of local water saturation also can be found in layers 1, 4, and 7. 

Furthermore, the local water dynamics within the majority of GDLs exhibit frequent fluctuations in Layer 13, attributed to droplet breakthrough and detachment in the central region of the GDL. In contrast to the preceding layers, the 13th layer demonstrates a higher saturation level in S-type GDLs and H2315 GDLs. An augmentation in fiber curvature results in a decline in water saturation in this layer, particularly evident in the transition from S-type GDLs to curved-fiber GDLs. Within the breakthrough region, water saturation primarily depends on the in-plane cross-section size of the fingering flow, as illustrated in Fig. \ref{GDLTotalSaDifferenceWithPores}, where the pore size distribution governs fingering flow size and length. The results also indicate that large pores in H2315 GDL and S-type GDLs form significant breakthrough pathways, while an abundance of small pores in curved-fiber GDLs increases the likelihood of smaller breakthrough pathways. Certain GDLs, such as C1-GDL2, C2-GDL1, and C2-GDL1, exhibit relatively stable saturation. 

\begin{figure}[H]
    \centering
    {\includegraphics[width = 0.9\textwidth]{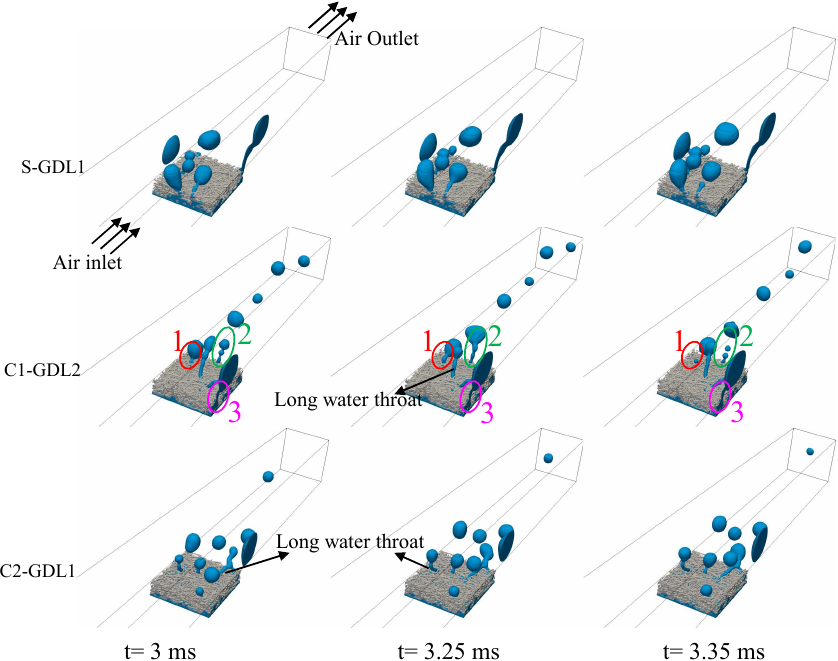}}
    \caption{Water behavior in three GCs connected with S-GDL1, C1-GDL2, and C1-GDL1, respectively. Based on C1-GDL2, three different breakthrough flow types in the GCs are labeled with 1 (red), 2 (green), and 3 (pink).}
    \label{WaterThroat}
\end{figure}

To explain the reasons behind these differences in oscillation degrees, Fig. \ref{WaterThroat} presents water flow within three interconnected GCs at 3 ms, 3.25 ms, and 3.35 ms. Based on water flow within the GC connected to C1-GDL2, the breakthrough flow is categorized into three types, labeled 1 (red), 2 (green), and 3 (pink). In Type-1 flow, droplets detach close to the GDL with a larger size, exerting a more substantial influence on GDL water. Type-2 flow features an elongated water throat and smaller droplets, resulting in a lower impact on GDL water. Type-3 flow adheres to the GC side wall, accumulating water without significant influence on GDL water. Consequently, a higher prevalence of Type-2 and Type-3 flow corresponds to reduced water oscillation in the top region of the GDL. The GC connected to C2-GDL1 also exhibits some Type-2 flow, correlating with smaller oscillations in GDL Layer 13, as depicted in Fig. \ref{TotalLayerSa}. In contrast, the water flow in the GC associated with S-GDL1 is characterized by a higher occurrence of Type-1 flow, explaining the frequent and pronounced oscillations. Following droplet detachment from GDL breakthrough water, a discernible decreasing trend in layer water saturation in the 13th layer is observed around 3 ms in Fig. \ref{TotalLayerSa}. Besides, the bigger droplets in the GCs move slowly compared with the small ones in the GCs connected with curved-fiber GDLs.

\section{Conclusions}
In this study, a rod periodic surface model stochastically reconstructs three Gas Diffusion Layer (GDL) types with varying fiber curvatures, using essential structural parameters from a physical Fredenberg H2315 GDL sample. Pore network and two-phase flow simulation models are employed to explore differences between reconstructed and physical GDLs, considering the impact of fiber curvature for the first time. The relationship between pore network parameters (pore diameter, throat diameter, coordination number) and GDL water saturation and capillary pressure is investigated. Although virtual GDLs maintain similar bulk and layer porosity, as well as fiber diameter, to the physical GDL, both simulation types reveal visible differences. H2315 GDL exhibits higher frequency in pore and throat diameter, and coordination number, and peak frequencies located in smaller sizes of these parameters. Despite the H2315 GDL having a more complex and better pore connection, it shows lower GDL water saturation and capillary pressure than virtual GDLs. Compared to all curved-fiber GDLs, straight-fiber GDLs, like S-GDL1, show potential similarity to H2315 GDL, but with a larger uncertain range in GDL water saturation and capillary pressure. Increasing fiber curvature leads to a noticeable rise in the amount of pore, throat, and coordination number. Moreover, capillary pressure shows an increase with fiber curvature. The increase in fiber curvature enhances complexity and pore connection level but contributes to larger capillary pressure due to the increased number of smaller pores. Long-term slow increase trends in GDL water saturation after water breakthrough becomes apparent with increasing fiber curvature. Water flow varies significantly from the GDL bottom to the top, and the breakthrough region experiences local water saturation oscillation during droplet attachment and detachment at the GDL/GC interface, indicating inherent instability. Three types of breakthrough flow are explored based on droplet detachment distance to the interface: near the interface, upon the interface, and no detachment on the side walls. Channels with straight-fiber GDLs show a higher occurrence of large droplets detaching close to the interface, resulting in a slower drainage process. The impact of droplet detachment near the interface on GDL top layer water oscillation is more pronounced than detachment upon the interface and attached flow on the GC side walls.

Future studies could extend to investigate water behavior in gas channels, such as pressure drop and water accumulation. Additionally, stochastic reconstruction could be improved by considering other GDL properties like tortuosity, permeability, and anisotropy for a more reliable and realistic base for further GDL optimization.

\section*{Declaration of Competing Interest}
The authors declare that they have no known competing financial interests or personal relationships that could have appeared to influence the work reported in this paper.

\section*{Acknowledgement}
The authors are sincerely grateful to Steven B. Beale for providing the image-based Fredeunberg H2315 GDL reconstruction used in this study. The authors also wish to thank the Chinese Scholarship Council for generous financial support, through grant number 202006070174. Additionally, the authors acknowledge the Swedish National Infrastructure for Computing (SNIC) for providing access to valuable computer resources under grant agreement no. 2018-05973.

\section{Model validation}\label{Appendix II}

\begin{table}[H]
\small
\centering
\caption{Parameter comparison between the present study and previous study \cite{niblett2020two}.}
\begin{tabular}{p{4.3cm}p{5.2cm}p{5cm}}
\hline
Parameters & Niblett et al. \cite{niblett2020two} & Present validation study \\ \hline
GDL Size & 786 $\mu m \times$ 848 $\mu m \times$ 165 $\mu m$ & 800 $\mu m \times$ 800 $\mu m \times$ 165 $\mu m$ \\
GC Size & 800 $\mu m \times$ - $\mu m \times$ 300 $\mu m$ & 800 $\mu m \times$ 800 $\mu m \times$ 300 $\mu m$ \\
GDL inlet & 500 $\mu m$ diameter, 50 $\mu m$ thickness & Same \\
GDL fiber contact angle & 139$^\circ$ & Same \\
GC wall contact angle & 56$^\circ$ & Same \\
Operation temperature & 80$^\circ$C & Same \\
Porosity & - & 0.88$^a$ \\
Fiber diameter & - & 9 $\mu m$ \cite{hinebaugh2017stochastic}  \\
Water inlet velocity & 0.0287 m/s & Same \\
Air inlet velocity & 15 m/s & Same \\ \hline
\multicolumn{3}{p{15cm}}{$^a$: \url{https://www.fuelcellstore.com/spec-sheets/SGL-GDL_24-25.pdf}} \\
\end{tabular}
\label{Modvalmy_label}
\end{table}

\section{Supplementary data}
A supplementary video is provided to show more details about the dynamic process of liquid water transport process through the physical GDL H2315 and three representative GDLs with different fiber curvature.

\bibliography{Reference}

\end{document}